\documentclass[11pt,technote,onecolumn]{IEEEtran}
\usepackage{times}
\usepackage{epsfig}
\usepackage{graphicx}
\usepackage{amsmath}
\usepackage{amssymb}
\usepackage{bm}
\usepackage{multirow}
\usepackage{color}
\usepackage[noend]{algpseudocode}
\usepackage{algorithmicx,algorithm}
\usepackage{array}
\usepackage{CJK}
\ifCLASSINFOpdf
\else
\fi
\begin{document}
%
\title{AMP-Net: Denoising based Deep Unfolding for Compressive Image Sensing}
%
%
%

\author{Zhonghao Zhang, Yipeng Liu ~\IEEEmembership{Senior Member,~IEEE}, Jiani Liu, Fei Wen, Ce Zhu ~\IEEEmembership{Fellow,~IEEE}
	\thanks{This research is supported in part by the National Natural Science Foundation of China (NSFC, No. 62020106011, No. U19A2052), in part by Sichuan Science and Technology Program (No. 2019YFH008). The corresponding author is Yipeng Liu.}
	\thanks{Zhonghao Zhang, Yipeng Liu, Jiani Liu and Ce Zhu are with School of Information and Communication Engineering, University of Electronic Science and Technology of China (UESTC), Chengdu 611731, China. (email: yipengliu@uestc.edu.cn).}
	\thanks{Fei Wen is with the Department of Electronic Engineering, Shanghai
	Jiao Tong University, Shanghai 200240, China.
}}

%
%

\markboth{JOurnal Name,~Vol.~XXX, No.~XXX, MOnth~Year}%
{Shell \MakeLowercase{\textit{et al.}}: Bare Demo of IEEEtran.cls for IEEE Journals}
%



\maketitle

\begin{abstract}
Most compressive sensing (CS) reconstruction methods can be divided into two categories, i.e. model-based methods and classical deep network methods. By unfolding the iterative optimization algorithm for model-based methods onto networks, deep unfolding methods have the good interpretation of model-based methods and the high speed of classical deep network methods.
In this paper, to solve the visual image CS problem, we propose a deep unfolding model dubbed AMP-Net. Rather than learning regularization terms, it is established by unfolding the iterative denoising process of the well-known approximate message passing algorithm. Furthermore, AMP-Net integrates deblocking modules in order to eliminate the blocking artifacts that usually appear in CS of visual images. 
In addition, the sampling matrix is jointly trained with other network parameters to enhance the reconstruction performance. 
Experimental results show that the proposed AMP-Net has better reconstruction accuracy than other state-of-the-art methods with high reconstruction speed and a small number of network parameters.
\end{abstract}


%
\IEEEpeerreviewmaketitle

\section{Introduction}
\label{introduction}
%
%
%
%

\IEEEPARstart{C}{ompressive} sensing (CS) requires much fewer measurements than the classical Nyquist sampling to reconstruct a signal \cite{candes2008introduction}. It has been applied in a series of imaging applications, including single-pixel camera~\cite{duarte2008single}, magnetic resonance imaging (MRI)~\cite{liu2017hybrid}, and snapshot compressive imaging (SCI)~\cite{ma2019deep}.

CS performs fast imaging by sampling few measurements, i. e. $ \mathbf{y} = \mathbf{A x} $, where $\mathbf{x} \in {\mathbb{R}^N}$ is the original signal, $\mathbf{y} \in  {\mathbb{R}^M}$ consists of the samples,  $\mathbf{A}\in {\mathbb{R}^{M \times N}}$ is the sampling matrix with $ M < N $. The image recovery from compressive samples is to solve an under-determined linear inverse system, and the corresponding optimization model can be formulated as follows: 

\begin{align}
\mathop {\min }\bm{\limits_\mathbf{x}} \mathfrak{R}(\mathbf{x})\,{\text{,~~s}}{\text{.~t}}{\text{.~~}}\mathbf{y} = \mathbf{A x},
\end{align}
where  $\mathfrak{R}(\mathbf{x})$ is the regularization term. 

To solve this optimization problem, {model-based} recovery methods exploit some data structures by employing structure-inducing regularizers \cite{baraniuk2010model}, such as sparsity in some transformation domains \cite{mallat1999wavelet, elad2010sparse, nam2013cosparse,Liu2020Smooth}, low rank \cite{cai2010singular, long2019low, liu2019image, liu2019low, Liu0Low, Liu2020Lowrank}, and so on \cite{liu2013multi, metzler2016denoising, zhang2017learning, ulyanov2018deep,Wu2019deep}. A number of non-linear iterative algorithms can be used to solve these optimization problems \cite{tropp2010computational, candes2010matrix, long2019low}, such as sparse Bayesian learning \cite{wipf2004sparse}, orthogonal matching pursuit (OMP) \cite{tropp2007signal}, fast iterative shrinkage-thresholding algorithm (FISTA) \cite{beck2009fast}, approximate message passing (AMP) \cite{donoho2009message}, etc. These methods usually have theoretical guarantees.

In these years, some {deep-learning-based} methods have been developed for the image recovery problem \cite{ravishankar2019image}.  Classical deep networks directly map the compressive samples as input to the {estimations} as output~\cite{dong2015image, mousavi2015deep, kulkarni2016reconnet, shi2019image, shi2019scalable}. These networks consist of stacked non-linear operational layers, such as autoencoders~\cite{mousavi2015deep}, convolutional neural networks (CNNs)~\cite{krizhevsky2012imagenet}, generative adversarial networks (GANs)~\cite{goodfellow2014generative}, etc. Their parameters can be trained by the well-known {backpropagation} algorithm. Compared with the iterative optimization algorithms, these classical deep neural networks can quickly reconstruct images. Nevertheless,  due to the black box characteristic of these models, there is no good interpretation and theoretical guarantee. {These} completely data-driven end-to-end manners may have {risks} for some undesired effects \cite{huang2018some}.  Therefore, it can be beneficial to integrate the prior knowledge and the structure of the operators. 

Motivated by the iterative algorithm that processes data step by step, deep unfolding maps iterative restoration algorithms onto deep neural networks  \cite{gregor2010learning, zhang2018ista, sun2016deep, dong2018denoising}. It tries to {make a compromise} between the iterative recovery methods and network methods, and enjoys a good balance between reconstruction speed and interpretation. Deep unfolding methods are first developed to solve the sparse linear inverse problem and the unfolded algorithms include ISTA~\cite{gregor2010learning, chen2018theoretical}, AMP~\cite{borgerding2017amp} and iterative hard thresholding (IHT) algorithm~\cite{wang2016learning}. However, it is usually necessary for these methods to obtain additional pre-trained dictionaries to recover images.

\begin{figure*}
	\setlength{\abovecaptionskip}{0pt}
	\begin{center}
		\includegraphics{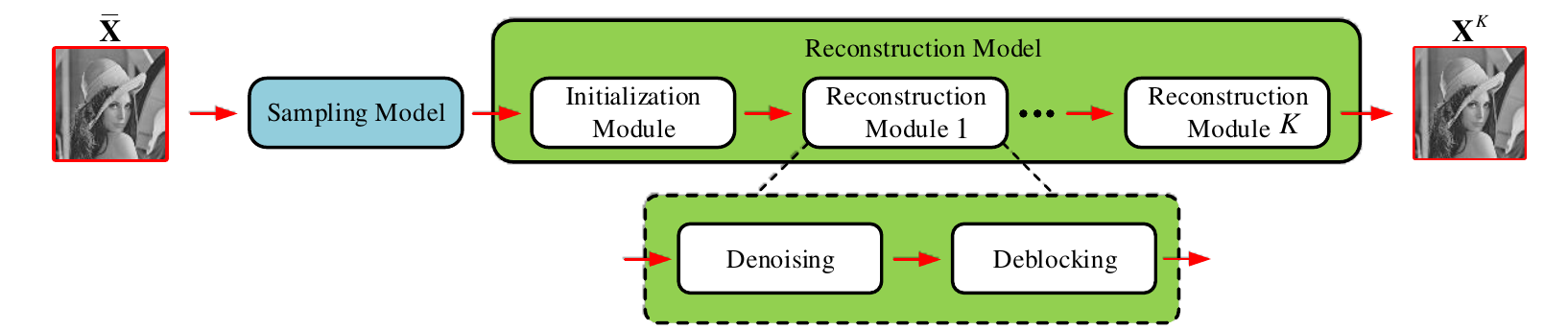}
	\end{center}
	\caption{The framework of AMP-Net. $\bar{\mathbf{X}}$ denotes the original image and ${{\mathbf{X}}^K}$ denotes the reconstructed image.}
	\label{fig:frame}
	\vspace{-0.3cm}
\end{figure*}

In recent years, some non-linear iterative algorithms are unfolded to solve image CS problems. For example, ADMM {is unfolded for MRI ~\cite{sun2016deep} and SCI~\cite{ma2019deep}}, and {the} primal-dual algorithm is developed to reconstruct CT~\cite{adler2018learned}. {AMP~\cite{metzler2017learned}} and ISTA~\cite{zhang2018ista} are exploited to reconstruct visual images.
And the half-quadratic splitting (HQS) algorithm~\cite{dong2018denoising} and the linear inverse operation~\cite{gilton2019neumann} are developed to solve common image inverse problems.
Some methods~\cite{sun2016deep,ma2019deep,zhang2018ista}  design deep unfolding models by pre-assuming specific regularizers, such as $\ell_1$ norm for sparsity in some basis. Carefully designed regularization terms are intuitive, but may ignore other image intrinsic structures. It has been proved that combining multiple kinds of image prior {knowledge} can get better reconstruction results than single one~\cite{liu2019image, liu2019low, lingala2011accelerated}. To integrate more information, some methods use deep networks to model an appropriate regularizer \cite{zhang2018ista,adler2018learned,dong2018denoising,gilton2019neumann}. 
But due to the uncertainty and unrepresentability of the regularizer,  it is difficult to learn the regularizer directly. Instead, some information related to the regularizer {is} learned, such as denoising prior~\cite{dong2018denoising}, gradient~\cite{diamond2017unrolled}, inverse information~\cite{gilton2019neumann}. 
However, to learn this information, classical deep neural networks usually contain a large {number} of parameters, which makes the application of related models limited.
In this paper, we establish the deep unfolding model by unfolding the iterative denoising process from the denoising perspective of the AMP algorithm. 

The AMP algorithm interprets a classical linear operation ${\mathbf{A}^{\operatorname{T}}}(\mathbf{y} - \mathbf{Ax'} ) + \mathbf{x'}$ as the sum of the original data and a noise term~\cite{donoho2009message}, where $\mathbf{x}' \in \mathbb{R}^{N}$ is an estimate of $\mathbf{x}$. To obtain the original data, related non-linear operations in each iteration can be regarded as a series of denoising processes which depend on different image priors. In this paper, we call this interpretation the denoising perspective of the AMP algorithm. {Based on this theory, ~\cite{metzler2017learned} employs CNNs for denoising or approximating the bias named Onsager correction term. However, the denoiser is designed without considering the sampling matrix. In this paper, we analyze the noise term in more detail, and {combine} CNNs and the sampling matrix $\mathbf{A}$ to fit the noise term jointly. Furthermore, based on that, a control parameter can be introduced to enhance the flexibility of the reconstruction.} Such a strategy makes the representation of the image prior more flexible and intuitive with fewer demanded parameters.

{Besides, in CS systems, some methods~\cite{li2013efficient,mousavi2015deep,kulkarni2016reconnet,zhang2018ista} sample and reconstruct images block-by-block. This strategy brings {less burden} to {the hardware}.}
However, for these block-based methods, additional deblocking operations must be applied to eliminate blocking {artifacts}. {Some methods~\cite{mousavi2015deep,kulkarni2016reconnet} use artificial deblockers to erase the {artifacts}, such as BM3D.} Some classical deep networks {perform image reconstruction} and deblocking at the same time ~\cite{shi2019image,shi2019scalable}. {In this paper, we propose a trainable deblocking module, which not only can be used in AMP-Net, but also can be plugged in other deep learning methods.}

In addition,  {recently the sampling matrix optimization is employed in some {model-based} methods and classical deep neural networks \cite{Wu2019deep,kulkarni2016reconnet,shi2019image,iliadis2020deepbinarymask,2018Deep}, which are designed for image CS\cite{Wu2019deep,kulkarni2016reconnet,shi2019image,2018Deep} and video CS\cite{iliadis2020deepbinarymask}.} In such a strategy, the sampling matrix is trained with other parameters using gradient descent way related algorithms~\cite{kingma2014adam}. {By limiting the value of the sampling matrix, different forms of sampling {matrices} can be obtained, including floating-point matrix~\cite{Wu2019deep,kulkarni2016reconnet,shi2019image}, sparse matrix~\cite{2018Deep} and binarized matrix~\cite{shi2019image,iliadis2020deepbinarymask}. In this paper, we focus on the trained floating-point matrix.} Since the sampling matrix plays an important role in both sampling and reconstruction in most deep unfolding methods~\cite{zhang2018ista, dong2018denoising, gilton2019neumann}, an appropriate sampling matrix may effectively improve the reconstruction performance. In this work, we jointly train the sampling matrix with other parameters of the designed deep unfolding model.

{In general,} in this paper, we propose a novel deep unfolding model dubbed AMP-Net to solve the  visual image CS problem. Fig. \ref{fig:frame} illustrates the framework of AMP-Net. AMP-Net is composed of a sampling model and a reconstruction model.
In the sampling model, images are measured block-by-block using the same sampling matrix. The reconstruction model is established by unfolding the iterative denoising process, which is inspired {by} the denoising perspective of the AMP algorithm. The reconstruction model is an unfolding form of the denoising process with $K$ iterations. It consists of an initialization module and $K$ reconstruction modules.
The initialization module is used to generate a reasonable initial estimation.  Each reconstruction module {contains} a denoising module and a deblocking module. The denoising module processes each image block, and {the} deblocking module is for the whole image. Experimental results demonstrate that the proposed method outperforms the state-of-the-art ones in terms of reconstruction accuracy and computational complexity. 

The main contributions of this paper can be summarized as follows:
\begin{itemize}
	\item We develop a novel deep unfolding model named AMP-Net which is inspired by the denoising perspective of the AMP algorithm. 
	{With a detailed analysis of the noise term, we {integrate} CNNs and the sampling matrix for accurate noise estimation.}
	
	\item 
	To eliminate the blocking {artifacts}, a trainable deblocking module is designed following the denoising module.
	To the best of our knowledge, this is the first work {which learns deblockers and integrates them into a deep unfolding model. And experimental results {reveal} that this strategy also benefits other deep learning methods.}
	
	\item  {
	
	 In AMP-Net, the sampling matrix is trained with other parameters jointly due to its contributions in both sampling and reconstruction. In this way, the produced optimized matrix is data-driven and provides  performance improvement not only for AMP-Net but for other reconstruction methods.
}
	
\end{itemize}

The paper is organized as follows. {Section \ref{background} introduces the denoising perspective of the AMP algorithm and some related works.} Section \ref{AMP_Net} describes  AMP-Net in detail. Section \ref{results} is the experimental results.  And in Section \ref{conclusion}, we conclude this paper.

\section{Background}
\label{background}
In this section, the denoising perspective of the AMP algorithm is introduced and some works related to AMP-Net are presented.

\subsection{The Denoising Perspective of the AMP Algorithm}

The AMP algorithm analyzes a classical scheme of iterative non-linear algorithms~\cite{beck2009fast, donoho2009message} as follows: 
\begin{align}
\label{for:denoise_before}
{\mathbf{z}^{k-1}} &= \mathbf{y} - \mathbf{A}{\mathbf{x}^{k-1}},
\\\label{equal4}{\mathbf{x}^{k}} = {\mathfrak{T}_k}&({\mathbf{A}^{\operatorname{T}}}{\mathbf{z}^{k-1}} + {\mathbf{x}^{k-1}}),
\end{align}
where ${\mathbf{A}^{\operatorname{T}}}$ is the transpose of sampling matrix $\mathbf{A}$, ${\mathfrak{T} _k}(\cdot )$ is the non-linear function, and $k$ is the number of the iteration. If the initialized input and the original data are defined as ${\mathbf{x}^0}$ and ${\bar{\mathbf{x}}}$, then we can have
\begin{align}
\label{equal5}
{\mathbf{A}^{\operatorname{T}}}{\mathbf{z}^0} + {\mathbf{x}^0} = {\bar{\mathbf{x}}} + ({\mathbf{A}^{\operatorname{T}}}\mathbf{A} - \mathbf{I})  ({\bar{\mathbf{x}}} - {\mathbf{x}^0}),
\end{align}
where $\mathbf{I}$ is the identity matrix in the size of $N \times N$. The detailed derivation of (\ref{equal5}) can be found in Appendix \ref{derivation1}.
By extending (\ref{equal5}) to the $k$-th iteration, we can get
\begin{align}
\label{equal6}
{\mathbf{A}^{\operatorname{T}}}{\mathbf{z}^{k-1}} + {\mathbf{x}^{k-1}} = {\bar{\mathbf{x}}} + ({\mathbf{A}^{\operatorname{T}}}\mathbf{A} - \mathbf{I})  ({\bar{\mathbf{x}}} - {\mathbf{x}^{k-1}}),
\end{align}
where the entries ${\mathbf{A}_{ij}} $ of the sampling matrix are independent and identically distributed as ${\mathbf{A}_{ij}}\sim\mathcal{N}(0,1/M)$, and $\mathcal{N} (\mu ,{\sigma ^2})$ denotes the Gaussian distribution with the mean value $\mu$ and the variance $\sigma ^2$. Under this assumption, it can be proved that $({\mathbf{A}^{\operatorname{T}}}\mathbf{A} -\mathbf{I}) (\bar{\mathbf{x}} - {\mathbf{x}^{k-1}})$  is also a Gaussian distributed vector with the variance ${M^{- 1}}\left\|\bar{\mathbf{x}}-{\mathbf{x}^{k-1}}\right\|_2^2$ ~\cite{donoho2009message}. Then, (\ref{equal6}) can be reformulated into the sum of the original signal and a noise term as follows:
\begin{align}
{\mathbf{A}^{\operatorname{T}}}{\mathbf{z}^{k-1}} + {\mathbf{x}^{k-1}} = {\bar{\mathbf{x}} } + \mathbf{e},
\end{align}
where $\mathbf{e}=({\mathbf{A}^{\operatorname{T}}}\mathbf{A} - \mathbf{I}) (\bar{\mathbf{x}} - {\mathbf{x}^{k-1}})$  denotes the noise term.

The AMP algorithm interprets the non-linear function $\mathfrak{T}_k(\cdot )$ as a denoising function varying with different signal priors. 
For example, if signals are assumed to be sparse without value limitation, $\mathfrak{T}_k(\cdot )$  can be the soft thresholding function~\cite{beck2009fast}. 
In this paper, we call such interpretation as the denoising perspective of the AMP algorithm.

Significantly, it is worth noting that $\mathbf{e}$ has no relation with the regularizer in {model-based} methods. It is only affected by the original image and the input of each iteration. Therefore, it can be fitted by supervised learning, while the image prior can be learned in the noise term learning process. Such a strategy makes the representation of the image prior more flexible and intuitive with no need to figure out the specific form of the regularization term.
And it can be noticed that if $\left\| {\bar{\mathbf{x}} - {\mathbf{x}^{k-1}}} \right\|_2^2$  decreases in each iteration, the Euclidean distance between the reconstruction result and $\bar{\mathbf{x}}$ would get smaller  as the iteration number increases.


\subsection{Related Works}
\label{related}
{Model-based} methods are based on regularization terms inspired by image priors. Sparsity in some transformation domains, such as DCT~\cite{mallat1999wavelet}, wavelet~\cite{mun2009block} and gradient domain~\cite{li2013efficient}, has been exploited to reconstruct visual images. For example, Li et al.~\cite{li2013efficient} used a second order total variation (TV) regularizer to build an optimization problem and applied the augmented Lagrangian method to recover each image block. However, the fixed domain may result in poor performance. To improve the reconstruction performance, some elaborate priors have been exploited, such as denoising prior~\cite{metzler2016denoising, zhang2017learning} and network prior~\cite{ulyanov2018deep, Wu2019deep}. Specifically, Metzler et al.~\cite{metzler2016denoising} combined the BM3D denoiser with the AMP algorithm to develop a new framework named D-AMP for image reconstruction. And Wu et al.~\cite{Wu2019deep} employed a pre-trained deep neural network to represent the image prior and applied three times of gradient descend to reconstruct images. In this framework namely DCS, the sampling matrix is pre-trained.

For classical deep network methods, Mousavi et al.~\cite{mousavi2015deep} designed
a stacked denoising autoencoder (SDA) to reconstruct images
from sampled data for the first time. After that, Lohit et al.~\cite{kulkarni2016reconnet} proposed a {CNN-based} model namely ReconNet to improve the reconstruction performance of visual images, {where a popular denoiser called BM3D is employed for image deblocking in postprocessing.}
Shi et al.~\cite{shi2019image} proposed a novel sampling-reconstruction framework named CSNet based on residual CNNs. {Although the sampling matrix can be jointly trained with the reconstruction model in these three methods, it is not integrated into the reconstruction model. For example, CSNet directly replaces the sampling model with a convolutional layer, which is not considered in the following reconstruction model.  However, in AMP-Net, the sampling matrix plays an important role in the reconstruction model. And by training the sampling matrix, while producing better measurements, the performance of the reconstruction model can be further improved.}

Deep unfolding methods combine the advantages of {model-based} methods and classical deep network methods. For example, the popular algorithm ISTA is unfolded as ISTA-Net~\cite{zhang2018ista}, which is commonly used to solve optimization problems with a sparsity-inducing regularizer. It {employs} CNNs to learn the appropriate transformation operations and trainable soft thresholding functions to reflect the sparsity of data.
Gilton et al.~\cite{gilton2019neumann} developed a deep unfolding model named Neumann Network (NN) for common image inverse problems by truncating Neumann series with a data-driven non-linear regularizer.

{Similar to AMP-Net, three methods namely DPDNN, LDIT and LDAMP apply denoising priors. Dong et al.~\cite{dong2018denoising} designed a deep unfolding model dubbed DPDNN to solve common inverse problems, which unfolds the denoising process inspired by HQS algorithm. In each reconstruction module of DPDNN, the deep neural network aims to learn the information of the regularizer, and two parameters are introduced to control the step size and the trade-off between the regularizer and the data fitting term. However, AMP-Net applies deep neural networks for noise approximation and introduces only one control parameter.
	
Metzler et al.~\cite{metzler2017learned} proposed LDIT and LDAMP inspired {by} the denoising-based iterative thresholding (DIT) algorithm and the AMP algorithm respectively. And different from LDIT, LDAMP introduces an addictive bias term named Onsager correction term before denoising.
The main difference between the proposed AMP-Net and LDIT or LDAMP is the method for noise estimation.  LDIT and AMP-Net both try to estimate the noise term which is formulated as $({\mathbf{A}^{\operatorname{T}}}\mathbf{A} - \mathbf{I})  (\bar{\mathbf{x}} - \mathbf{x}^{k-1})$. However, they employ different estimation strategies. In detail, LDIT employs a CNN to estimate the entire noise term directly, but AMP-Net uses a CNN to fit $(\bar{\mathbf{x}} - \mathbf{x}^{k-1})$ only and then multiplies the sampling matrix to estimate the noise. In addition, a trainable parameter $\alpha _k$ is introduced to improve the flexibility of AMP-Net.	
}

\section{AMP-Net}
\label{AMP_Net}
In this section, we will illustrate the detail of AMP-Net.
As shown in Fig. \ref{fig:frame}, AMP-Net is composed of a sampling model and a reconstruction model which contains an initialization module and a series of stacked reconstruction modules. 

Note that we mainly focus on single-channel images in this study while colorful images can be sampled and recovered channel by channel.

\subsection{Sampling Model}

The sampling model samples images block-by-block and is similar to the one in CSNet~\cite{shi2019image}. However, it is not composed of convolutional kernels but a simple sampling matrix due to its relevance to the reconstruction process.

\begin{figure}
	\setlength{\abovecaptionskip}{0.cm}
	\begin{center}
		\includegraphics{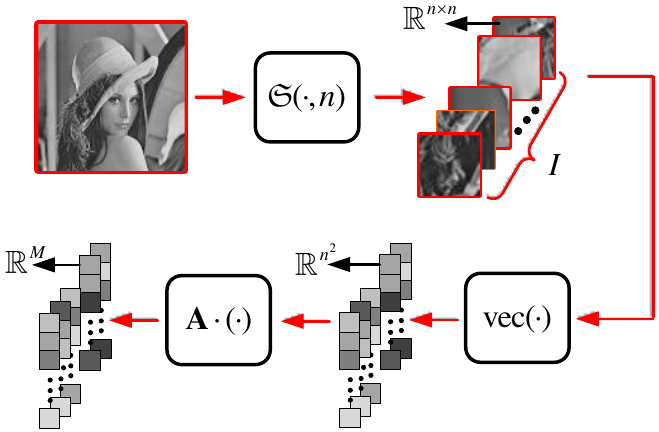}
	\end{center}
	\caption{The sampling process of the sampling model.}
	\label{fig:samp}
	\vspace{-3mm}
\end{figure}

To demonstrate the sampling process, we denote $\mathfrak{S}(\mathbf{X},n)$ as a splitting function which divides a single-channel image as $\mathbf{X} \in {\mathbb{R}^{L \times P}}$ into a series of {non-overlapping} image blocks. Each image block is denoted as  ${\mathbf{X}_{i}} \in {\mathbb{R}^{n \times n}}$ , where $i \in \{1,2, \cdots ,I\} $ and $L \cdot P = I \cdot n^2$. And $\operatorname{vec}( \cdot )$ is defined as a vectorization function which vectorize an image block to a vector, and  satisfies $\operatorname{vec}({\mathbf{X}_{i}}) \in {\mathbb{R}^{{n^2}}}$ and $\operatorname{vec}(\mathfrak{S}(\mathbf{X},n)) \in {\mathbb{R}^{{n^2} \times I}}$.  The sampling process of the sampling model is shown in Fig. \ref{fig:samp} and can be expressed as
\begin{align}
\label{equal8}
\mathbf{Y} = \mathbf{A}  \operatorname{vec}(\mathfrak{S}(\mathbf{X},n)),
\end{align} 
where $\mathbf{A} \in {\mathbb{R}^{M \times {n^2}}}$ is the sampling matrix for image blocks  and  $\mathbf{Y} \in {\mathbb{R}^{M \times I}}$ is the measurement. Each column of $\mathbf{Y}$ is the vectorized measurement of an image block.

Moreover, in order to further improve the performance of AMP-Net, $\mathbf{A}$ is simultaneously trained with other parameters for its contributions in both sampling and reconstruction. And the derivation of its gradient for updating can refer to Appendix \ref{derivation3}. 

\begin{figure}
	\setlength{\abovecaptionskip}{0.cm}
	\begin{center}
		\includegraphics{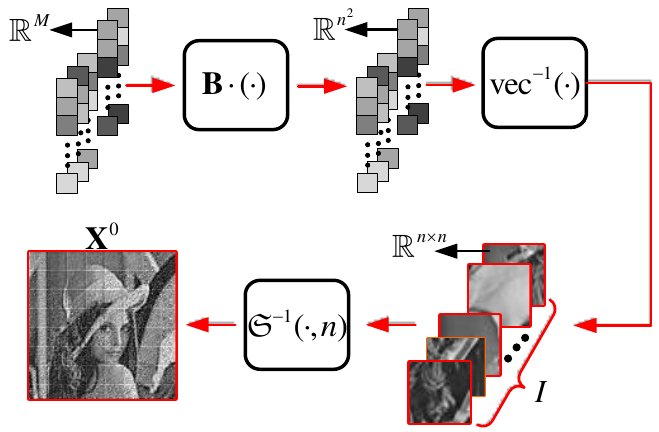}
	\end{center}
	\caption{The initialization process of the initialization module.}
	\label{fig:init}
	\vspace{-3mm}
\end{figure}

\subsection{Reconstruction Model}
\label{section:recon_model}
Due to the inspiration of the denoising perspective of the AMP algorithm, the reconstruction model is established by unfolding the iterative denoising process. This model is composed of an initialization module and a series of reconstruction modules. 
The initialization module is used to generate a reasonable initial estimation.
The subsequent reconstruction modules are derived by mapping the iterative denoising process onto a deep network. Each module stands for an iteration.
And each reconstruction module contains a denoising module and a deblocking module. 

\begin{figure*}
	\begin{center}
		\includegraphics{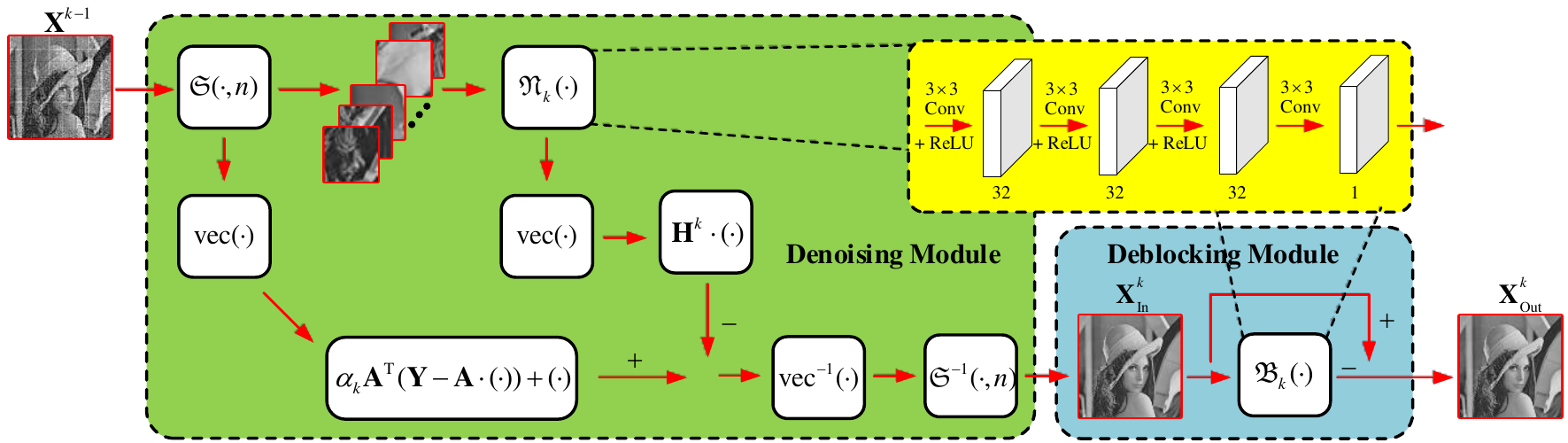}
	\end{center}
	\caption{The $k$-th reconstruction module of AMP-Net.}
	\label{fig:recon}
	\vspace{-3mm}
\end{figure*}

\emph{ Initialization module.} The image is initialized by linear operations of the observations. Fig. \ref{fig:init} shows the initialization process. ${\mathfrak{S}^{ - 1}}( \cdot ,n)$ is defined as a concatenation function which merges all the image blocks into a whole image and satisfies $\mathbf{X} = {\mathfrak{S}^{ - 1}}(\mathfrak{S}(\mathbf{X},n),n)$. Meanwhile,  ${\operatorname{vec}^{ - 1}}( \cdot )$ is a reshaping function which reshapes the vectorized image to its original shape and satisfies ${\mathbf{X}_{i}} = {\operatorname{vec}^{ - 1}}(\operatorname{vec}({\mathbf{X}_{i}}))$. The initialization process can be formulated as 
{\begin{align}
\label{equal9}
{\mathbf{X}^0} = {\mathfrak{S}^{ - 1}}({\operatorname{vec}^{ - 1}}(\mathbf{B}\mathbf{Y}),n),
\end{align}}
where $\mathbf{B} \in {\mathbb{R}^{{n^2} \times M}}$  is a trainable matrix for initialization and ${\mathbf{X}^0} \in {\mathbb{R}^{{L \times P}}}$ denotes the initialized image.

\emph{Denoising module.} The denoising module is designed to reconstruct each image block. By assuming that ${\mathbf{x}_{i}} = \operatorname{vec}({\mathbf{X}_{i}})$ and inspired by (\ref{equal6}), the original data can be obtained by calculating
\begin{align}
\label{equal10}
\bar{\mathbf{x}}_{i} = &{\mathbf{A}^{\operatorname{T}}}\mathbf{z}_{i}^{k-1} + \mathbf{x}_{i}^{k-1} \notag
\\ &- ({\mathbf{A}^{\operatorname{T}}}\mathbf{A} - \mathbf{I})  (\bar{\mathbf{x}}_{i} - \mathbf{x}_{i}^{k-1}).
\end{align}
If $\bar{\mathbf{x}}_{i}-\mathbf{x}_{i}^{k-1}$ is obtained, the reconstruction can be achieved by linear operations. To this end, we replace  $\bar{\mathbf{x}}_{i}-\mathbf{x}_{i}^{k-1}$ with a non-linear trainable function  ${\mathfrak{N}_k}( \cdot )$ and extend (\ref{equal10}) to an iterative version, which can be expressed as
\begin{align}
\label{equal11}
\mathbf{x}_{i}^{k} = &{\mathbf{A}^{\operatorname{T}}}\mathbf{z}_{i}^{k-1} + \mathbf{x}_{i}^{k-1} \notag
\\&- ({\mathbf{A}^{\operatorname{T}}}\mathbf{A} - \mathbf{I})  \operatorname{vec}({\mathfrak{N}_k}(\mathbf{X}_{i}^{k-1}))).
\end{align}
By regarding $({\mathbf{A}^{\operatorname{T}}}\mathbf{A}-\mathbf{I})  \operatorname{vec}({\mathfrak{N}_k}(\mathbf{X}_{i}^{k-1})))$ as the noise term, the reconstruction process in (\ref{equal11}) can be interpreted as the denoising process. 
{And because ${\mathfrak{N}_k}(\mathbf{X}_{i}^{k-1})$ is used to fit $\bar{\mathbf{X}}_{i}-\mathbf{X}_{i}^{k-1}$, the noise term is not limited to the Gaussian noise.}

Moreover, a trainable parameter ${\alpha _k}$ is introduced to enhance the flexibility of the reconstruction process. With ${\alpha _k}$, the iterative reconstruction process can be expressed as
\begin{align}
\label{equal12}
\mathbf{x}_{i}^{k} = &{\alpha _k}{\mathbf{A}^{\operatorname{T}}}\mathbf{z}_{i}^{k-1} + \mathbf{x}_{i}^{k-1}  \notag
\\&- ({\alpha _k}{\mathbf{A}^{\operatorname{T}}}\mathbf{A} - \mathbf{I})  \operatorname{vec}({\mathfrak{N}_k}(\mathbf{X}_{i}^{k-1})).
\end{align}
The detailed derivation of (\ref{equal12}) is similar to (\ref{equal11}) and can be found in  Appendix \ref{derivation2}.  Significantly, ${\alpha_k}$ is similar to the step size in other deep unfolding algorithms~\cite{chen2018theoretical,zhang2018ista}, but it is a parameter to control the noise and the reconstruction ability of the module in this paper. Therefore, ${\alpha_k}$ is named as the control parameter, and  (\ref{equal12}) would degenerate to (\ref{equal11}) when ${\alpha_k}$ equals 1.


In this paper, ${\mathfrak{N}_k}( \cdot )$ is designed to be a CNN. As shown in Fig. \ref{fig:recon},  ${\mathfrak{N}_k}( \cdot )$ is constructed of four convolutional layers, of which the first three layers with bias terms are followed by the Rectified Linear Unit (ReLU)~\cite{nair2010rectified} and the last convolutional layer has no bias term. The filter size of each convolutional layer is $3 \times 3$. To make sure that the output has the same size as the input, the padding size of each convolutional layer is set to be 1. The {numbers} of output channels {of} the four convolutional layers are 32, 32, 32 {and} 1, respectively.

By extending (\ref{equal12}) to the reconstruction of the whole image, the mathematical expression of the denoising module of the $k$-th reconstruction module is as follows:
\begin{align}
\label{equal13}
{\mathbf{X}^{k}} = &{\mathfrak{S}^{ - 1}}({\operatorname{vec}^{ - 1}}({\alpha _k}{\mathbf{A}^{\operatorname{T}}}{\mathbf{Z}^{k-1}} + \operatorname{vec}(\mathfrak{S}({\mathbf{X}^{k-1}},n))  \notag\\
&- \mathbf{H}^{k}  \operatorname{vec}({\mathfrak{N}_k}(\mathfrak{S}({\mathbf{X}^{k-1}},n)))),n), 
\end{align} 
where
\begin{align}
\mathbf{H}^{k}  = &{\alpha _k}{\mathbf{A}^{\operatorname{T}}}\mathbf{A} - \mathbf{I},\\
{\mathbf{Z}^{k-1}} = \mathbf{Y} - &\mathbf{A}  \operatorname{vec}(\mathfrak{S}({\mathbf{X}^{k-1}},n)).
\end{align} 
And Fig. \ref{fig:recon} provides a graphical illustration of the $k$-th reconstruction module of AMP-Net.

As we can see, $\mathbf{A}$ {plays} an important role in the reconstruction process (\ref{equal13}). If there is an appropriate sampling matrix, the reconstruction performance would get improved. This probably partly explains why the sampling matrix training strategy is effective for recovery.
{Furthermore, we emphasize that $\mathfrak{S}( \cdot ,n)$, ${\mathfrak{S}^{ - 1}}( \cdot ,n)$, $\operatorname{vec}( \cdot )$  and ${\operatorname{vec}^{ - 1}}( \cdot )$ do not introduce additional floating-point calculations.}

\emph{Deblocking module.} Reconstructing images block-by-block without overlapping may lead to a situation that additional deblocking operations must be carried out. {Some methods~\cite{mousavi2015deep,kulkarni2016reconnet} use hand-crafted denoisers to deblock images.} {However, artificial denoisers may ignore some semantic information that is helpful for deblocking. In this paper, a trainable deblocking module} is developed to eliminate the blocking {artifacts} and further improve the reconstruction performance.

\begin{table*}
	\setlength{\belowcaptionskip}{0pt}
	\setlength{\abovecaptionskip}{2pt}
	\begin{center}
		\footnotesize
		\caption{The comparison between AMP-Net and other state-of-the-art methods.}
		\begin{tabular}{|l|c|c|c|c|c|c|}
			\hline
			\multirow{2}*{{\bf Method}} & \multirow{2}*{\bf Trainablity} & \multirow{2}*{\bf Interpretation} &{\bf Fixed computational}   &{\bf Sampling matrix} &{\bf Deblocking} &{\bf Regularization} \\
			&&&{\bf complexity}&{\bf training}&{\bf operation}&{\bf term}\\
			\hline\hline
			TVAL3~\cite{li2013efficient} & & $ \bm\surd$ & &   &  & $ \bm\surd$\\
			\cline{2-7}
			D-AMP~\cite{metzler2016denoising} && $ \bm\surd$  &  & & & $\bm\surd$\\ 
			\cline{2-7}
			DCS~\cite{Wu2019deep} &$ \bm\surd$ & $\bm\surd$  & &  $\bm\surd$ &  &$\bm\surd$\\ 
			\cline{2-7}
			ReconNet~\cite{kulkarni2016reconnet} &$ \bm\surd$ &  &$\bm\surd$ & $\bm\surd$ & $\bm\surd$ & \\
			\cline{2-7}
			${\text{CSNe}}{{\text{t}^+}}$~\cite{shi2019image} &$ \bm\surd$ &  &$\bm\surd$ & $\bm\surd$ &$\bm\surd$  &\\
			\cline{2-7}
			${\text{ISTA-Ne}}{{\text{t}}^+}$~\cite{zhang2018ista} &$ \bm\surd$ & $\bm\surd$ &$\bm\surd$ & &  &$\bm\surd$ \\
			\cline{2-7}
			LDIT~\cite{metzler2017learned} &$ \bm\surd$ & $\bm\surd$ &$\bm\surd$ & & & \\
			\cline{2-7}
			LDAMP~\cite{metzler2017learned} &$ \bm\surd$ & $\bm\surd$ &$\bm\surd$ & & & \\
			\cline{2-7}
			DPDNN~\cite{dong2018denoising}&$ \bm\surd$ &$\bm\surd$&$\bm\surd$&&&$\bm\surd$\\
			\cline{2-7}
			NN~\cite{gilton2019neumann}&$ \bm\surd$ &$\bm\surd$&$\bm\surd$&&&$\bm\surd$\\
			\cline{2-7}
			AMP-Net &$ \bm\surd$ & $\bm\surd$ &$\bm\surd$  &  $\bm\surd$ & $\bm\surd$ &\\
			\hline
		\end{tabular}
		\label{tab:methods}
	\end{center}
	\vspace{-3mm}
\end{table*}

In detail, $\mathbf{X}^k$ can be further expressed as $\mathbf{X}^k=\bar{\mathbf{X}}+\mathbf{E}$ where $\mathbf{E} \in \mathbb{R}^{L \times P}$ {denotes the additive blocking {artifacts} from the $k$-th denoising module.} To map $\mathbf{X}^k$ to $\bar{\mathbf{X}}$, a non-linear learnable function ${\mathfrak{B}_k}( \cdot )$ is designed to fit $\mathbf{E}$. With {the} similar structure of ResNet~\cite{he2016deep}, the process of image deblocking in the $k$-th reconstruction module is illustrated in Fig. \ref{fig:recon}
and can be expressed as
\begin{align}
\mathbf{X}_{{\text{Out}}}^k = \mathbf{X}_{{\text{In}}}^k - {\mathfrak{B}_k}(\mathbf{X}_{{\text{In}}}^k),
\end{align} 
where $\mathbf{X}_{{\text{In}}}^k$ and $\mathbf{X}_{{\text{Out}}}^k$ denote the input and the output, respectively.  The input of ${\mathfrak{B}_k}( \cdot )$  is the whole concatenated image rather than each image block. In fact, ${\mathfrak{B}_k}( \cdot )$ is a CNN with the same structure as ${\mathfrak{N}_k}( \cdot )$ in this study but different parameter values. Significantly, this process can also be regarded as the further denoising of the image{,} and processing the whole image gives AMP-Net the potential of image deblocking.

\begin{algorithm}[t]
	\caption{The forward propagation of AMP-Net-$K$-BM} 
	\hspace*{0.02in} {\bf Input:} 
	$\bar{\mathbf{X}}$, $\mathbf{A}$, $\mathbf{B}$, $\mathbb{S}_{\alpha}$, $\mathbb{S}_{\mathbf{\Theta}}$, $\mathbb{S}_{\mathbf{\Omega}}$, $n$, $K$\\
	\hspace*{0.02in} {\bf Output:} 
	output $\mathbf{X}^{K}$\\
	\hspace*{0.02in} {\bf Sampling process:}
	\begin{algorithmic}[0]
		\State {\hspace{4pt} $\mathbf{Y} = \mathbf{A}  \operatorname{vec}(\mathfrak{S}(\bar{\mathbf{X}},n))$} 
	\end{algorithmic}
	\hspace*{0.02in} {\bf Reconstruction process:}
	\begin{algorithmic}[1]
		\State Set $k=0$
		\State ${\mathbf{X}^k} = {\mathfrak{S}^{ - 1}}({\operatorname{vec}^{ - 1}}(\mathbf{B}\mathbf{Y}))$
		\For{$k< K$} 
		\State $k \leftarrow k+1$
		\State $\mathbf{H}^{k}  = {\alpha_k}{\mathbf{A}^{\operatorname{T}}}\mathbf{A} - \mathbf{I}$
		\State ${\mathbf{Z}^{k-1}} = \mathbf{Y} - \mathbf{A}  \operatorname{vec}(\mathfrak{S}({\mathbf{X}^{k-1}},n))$
		\State {\setlength\abovedisplayskip{-12pt}
			\begin{align}
			\hspace{3pt}{\mathbf{X}^{k}} = &{\mathfrak{S}^{ - 1}}({\operatorname{vec}^{ - 1}}({\alpha _k}{\mathbf{A}^{\operatorname{T}}}{\mathbf{Z}^{k-1}} + \operatorname{vec}(\mathfrak{S}({\mathbf{X}^{k-1}},n))  \notag\\
			&- \mathbf{H}^{k} \operatorname{vec}({\mathfrak{N}_k}(\mathfrak{S}({\mathbf{X}^{k-1}},n)))),n) \notag 
			\end{align}}
		\State $\mathbf{X}^{k} = \mathbf{X}^{k} - {\mathfrak{B}_k}(\mathbf{X}^{k})$
		\EndFor
		\State \Return $\mathbf{X}^{k}$ 
	\end{algorithmic}

\end{algorithm}

\subsection{Loss Function}
{In this paper, for easy reference, AMP-Net-$K$ is named as the AMP-Net with $K$ denoising modules.
AMP-Net-$K$-B is AMP-Net-$K$ with deblocking modules and AMP-Net-$K$-M is AMP-Net-$K$ with the sampling matrix training strategy. And AMP-Net-$K$-BM is AMP-Net-$K$ with both of them.}

The trainable parameters of AMP-Net-$K$-BM  contain the measurement matrix $\mathbf{A}$,  initialization matrix $\mathbf{B}$, the control parameters $\mathbb{S}_{\alpha} = \{ {\alpha _1},{\alpha _2}, \cdots, {\alpha _K}\} $, all the trainable parameters in $\mathfrak{N}( \cdot )$ set as  $\mathbb{S}_{\mathbf{\Theta}}  = \{ {\mathbf{\Theta}_1},{\mathbf{\Theta} _2}, \cdots ,{\mathbf{\Theta} _K}\} $, and all the trainable parameters in $\mathfrak{B}( \cdot )$ set as $\mathbb{S}_{\mathbf{\Omega}} = \{ {\mathbf{\Omega} _1},{\mathbf{\Omega} _2}, \cdots ,{\mathbf{\Omega} _K}\} $. 
${\mathbf{\Theta} _k}$ denotes the trainable parameters of  ${\mathfrak{N}_k}( \cdot )$ and ${\mathbf{\Omega} _k}$ denotes the trainable parameters of  ${\mathfrak{B}_k}( \cdot )$. 
Algorithm 1 describes the forward propagation of AMP-Net-$K$-BM.

In this paper, we use mean square error (MSE) to describe the difference between the original image and the recovered image. Then the loss function of AMP-Net-$K$-BM can be formulated as
\begin{align}
\label{equ:loss}
\mathfrak{L}(\mathbf{A}, \mathbf{B}, \mathbb{S}_{\alpha}, \mathbb{S}_{\mathbf{\Theta}}, \mathbb{S}_{\mathbf{\Omega}} ) =
\frac{1}{{N_{\text{a}}{N_{\text{b}}}}}\sum\limits_{m = 1}^{{N_{\text{b}}}} {\left\| {\bar{\mathbf{X}}_m-\mathbf{X}_m^K} \right\|_2^2},
\end{align} 
where $\bar{\mathbf{X}}_m$ is the $m$-th original image in the training set, $N_{\text{a}}$ denotes the size of $\bar{\mathbf{X}}_m$ and ${N_{\text{b}}}$  denotes the size of the training set. {We emphasize that the forward propagation processes and the loss functions of AMP-Net-$K$, AMP-Net-$K$-B and AMP-Net-$K$-M can be easily derived from Algorithm 1 and (\ref{equ:loss}) by erasing some elements.}

\section{Experimental Results}
\label{results}
In this section, numerical experiments are performed on the visual image CS reconstruction.
{Firstly, the effectiveness of the unfolding strategy, deblocking module and sampling matrix learning strategy of AMP-net is validated individually in section IV-B, IV-C and IV-D. Then the performance of AMP-Net with blocking modules and the sampling matrix optimization is compared with some {state-of-the-art} reconstruction methods.}

\begin{table*}
	\setlength{\belowcaptionskip}{0pt}
	\setlength{\abovecaptionskip}{2pt}
	\begin{center}
		\footnotesize
		\caption{The test results of eight comparison methods and AMP-Net-6 and AMP-Net-9 on Set11. All methods apply no deblocking strategy and no sampling matrix strategy.}
		\begin{tabular}{|l|c|c|c|c|c|c|c|}
			\hline
			\multirow{2}*{{\bf Method}} & 50$\%$ & 40$\%$ &30$\%$ &25$\%$ &10$\%$ &4$\%$ &1$\%$\\
			\cline{2-8}
			& \multicolumn{7}{c|}{PSNR (dB)/SSIM}\\
			\hline\hline
			
			TVAL3 & 33.39/0.8157  &31.21/0.7531 & 29.00/0.6764 & 27.63/0.6238 & 22.45/0.3758 & 17.88/0.1997 & 14.90/0.0646 \\ 
			
			D-AMP& 37.34/0.8504  &35.22/0.8078 & 32.64/0.7544 & 31.62/0.7233 & 19.87/0.3757 & 11.28/0.0971 & 5.58/0.0034\\ 
			
			ReconNet & 32.12/0.9137 &30.59/0.8928 & 28.72/0.8517 & 28.04/0.8303 & 24.07/0.6958 & 21.00/0.5817 & 17.54/0.4426\\
			
			LDIT& 37.06/0.9626 &34.92/0.9470 & 32.69/0.9223 & 31.35/0.9042 & 25.56/0.7691 & 21.45/0.6075 & 17.58/0.4449\\
			LDAMP& 35.90/0.9531 &34.07/0.9383 & 32.01/0.9144 & 29.93/0.8783 & 24.94/0.7483 & 21.30/0.5985 & 17.51/0.4409\\
			DPDNN&35.85/0.9532&34.30/0.9411&32.06/0.9145&30.63/0.8924&24.53/0.7392&21.11/0.6029&17.59/{\bf0.4459}\\
			NN&31.41/0.8871&29.51/0.8523&27.64/0.8095&26.57/0.7842&22.99/0.6591&20.65/0.5525&{\bf17.67}/0.4324\\
			${\text{ISTA-Ne}}{{\text{t}}^+}$ & {\bf38.08}/{\bf0.9680} &{\bf35.93}/{\bf0.9537} & {\bf33.66}/{\bf0.9330} & {\bf32.27}/{\bf0.9167} & {\bf25.93}/{\bf0.7840}  & 21.14/0.5947 & 17.48/0.4403\\
			AMP-Net-6&37.48/0.9650&35.27/0.9495&32.96/0.9254&31.71/0.9090&25.72/0.7750&21.52/0.6100&17.53/0.4426\\
			AMP-Net-9&37.78/0.9667&35.63/0.9523&33.40/0.9307&32.05/0.9140&{\bf25.93}/0.7828&{\bf21.69}/{\bf0.6224}&17.58/0.4397\\
			\hline
		\end{tabular}
		\label{tab:set11}
	\end{center}
	\vspace{-3mm}
\end{table*}
\begin{table*}
	\setlength{\belowcaptionskip}{0pt}
	\setlength{\abovecaptionskip}{2pt}
	\begin{center}
		\footnotesize
		\caption{The test results of eight comparison methods and AMP-Net-6 and AMP-Net-9 on the test set of BSDS500. All methods apply no deblocking strategy and no sampling matrix strategy.}
		\begin{tabular}{|l|c|c|c|c|c|c|c|}
			\hline
			\multirow{2}*{{\bf Method}} & 50$\%$ & 40$\%$ &30$\%$ &25$\%$ &10$\%$ &4$\%$ &1$\%$  \\
			\cline{2-8}
			& \multicolumn{7}{c|}{PSNR (dB)/SSIM} \\
			\hline\hline
			
			ReconNet & 30.85/0.8949 &29.47/0.8647 & 27.95/0.8190 & 27.20/0.7914 & 23.98/0.6472 & 21.69/0.5557 & 18.96/0.4531\\
			LDIT & 34.27/0.9453 &32.23/0.9184 & 30.23/0.8799 & 29.10/0.8523 & 24.94/0.7040 & 22.04/0.5759 & 19.00/{\bf0.4564}\\
			LDAMP& 33.45/0.9359 &31.79/0.9116 & 29.89/0.8724 & 28.35/0.8297 & 24.61/0.6920 & 21.93/0.5721 & 18.94/0.4512\\
			DPDNN&33.56/0.9373&32.05/0.9164&29.98/0.8759&28.87/0.8491&24.37/0.6863&21.80/0.5716&18.97/0.4544\\
			NN&30.47/0.8882&28.84/0.8511&27.23/0.8037&26.42/0.7757&23.44/0.6443&21.49/0.5451&{\bf19.06}/0.4474\\
			${\text{ISTA-Ne}}{{\text{t}}^+}$ & {\bf34.92}/{\bf0.9510} &{\bf32.87}/{\bf0.9264} & {\bf30.77}/{\bf0.8901} & {\bf29.64}/{\bf0.8638} & 25.11/0.7124  & 21.82/0.5661 & 18.92/0.4529\\
			AMP-Net-6&34.55/0.9478&32.50/0.9223&30.43/0.8840&29.33/0.8574&25.02/0.7095&22.09/0.5779&18.99/0.4543\\
			AMP-Net-9&34.79/0.9504&32.75/0.9253&30.67/0.8887&29.53/0.8625&{\bf25.12}/{\bf0.7153}&{\bf22.22}/{\bf0.5851}&18.97/0.4541\\
			\hline
		\end{tabular}
		\label{tab:bsd500}
	\end{center}
	\vspace{-3mm}
\end{table*}

\subsection{Experimental settings}
All of our experiments are performed on two datasets: BSDS500~\cite{arbelaez2010contour} and Set11~\cite{kulkarni2016reconnet}. BSDS500 contains 500 colorful visual images which are divided into three parts: the training set (200 images), the validation set (100 images) and the test set (200 images). And Set11~\cite{kulkarni2016reconnet} contains 11 grey-scale images. We use the luminance components of BSDS500 for training, validation and testing, and use Set11 for testing. 

In this study, we set the size $n$ of each image block as 33. {Two training sets are generated for models with and without trainable deblocking operations respectively. (a) Training set 1: 448 sub-images with the size $99 \times 99$  are randomly extracted from the luminance component of each image in the training set of BSDS500~\cite{shi2019image}. (b) Training set 2: 977 images with the size $33 \times 33$ are randomly extracted from the luminance component of each image in the training set of BSDS500~\cite{zhang2018ista}.}
The validation set of BSDS500 is used to choose the best model for testing. And the test set of BSDS500 and Set11 are used for testing.
Peak Signal-to-Noise Ratio (PSNR) and Structural Similarity Index (SSIM) are used for evaluation. The higher PSNR and SSIM are, the better the models perform. The average PSNR on the validation set are calculated at the end of each training epoch, and the model with the highest PSNR is regarded as the best model for testing. 

Before training, $\mathbf{A}$ is initialized randomly as a Gaussian matrix and its rows are orthogonalized.
The control parameter $\alpha_k $ is initialized as 1, $\mathbf{B}$ is initialized as $\mathbf{A}^{\operatorname{T}}$ and other trainable parameters are initialized randomly. And we emphasize that all models use the same initialized sampling matrix when the CS ratio is the same.
The optimization algorithm employed for training is Adam~\cite{kingma2014adam}. 
{In this section, the numbers of the reconstruction module of AMP-Net are set to 2, 4, 6, or 9. All models are trained for 100 epochs with batch size 32 and learning rate 0.0001.}
All the experiments are implemented on a platform with an AMD Ryzen7 2700X CPU and an RTX2080Ti GPU.

{Methods compared with AMP-Net are TVAL3~\cite{li2013efficient}, D-AMP~\cite{metzler2016denoising}, DCS~\cite{Wu2019deep}, ReconNet~\cite{kulkarni2016reconnet},  ${\text{CSNe}}{{\text{t}}^+}$~\cite{shi2019image}, LDIT~\cite{metzler2017learned}, LDAMP~\cite{metzler2017learned}, DPDNN~\cite{dong2018denoising}, NN~\cite{gilton2019neumann}, ${\text{ISTA-Ne}}{{\text{t}}^+ }$~\cite{zhang2018ista}. TVAL3, D-AMP, DCS are {model-based} methods. ReconNet and ${\text{CSNe}}{{\text{t}}^+}$ are classical deep network methods.
	 LDIT, LDAMP, DPDNN, NN and ${\text{ISTA-Ne}}{{\text{t}}^+ }$ are deep unfolding methods.
	In this paper, all methods are modified accordingly to solve the block-based image CS problem.
	Table \ref{tab:methods} highlights the difference between these comparison methods and our method, in terms of whether to have interpretation, whether to have deblocking operations, whether to apply the sampling matrix training strategy and more.
	Significantly, since LDIT and LDAMP are also derived from {the} AMP algorithm, for fair comparison,
	their residual CNNs are designed to have the same number of layers as ${\mathfrak{N}_k}( \cdot )$ of AMP-Net in this paper.}

\begin{table*}
	\setlength{\abovedisplayskip}{3pt}
	\setlength{\belowdisplayskip}{3pt}
	\begin{center}
		\footnotesize
		\caption{The test results of models tested on the test set of Set11 with different deblocking strategies. No-D, BM3D and DM mean reconstructing with no deblocking strategy, BM3D denoiser and the deblocking module, respectively.}

		\begin{tabular}{|l|c|c|c|c|c|c|}
			\hline
			\multirow{3}*{{\bf Method}}&\multicolumn{3}{|c|}{30\%}&\multicolumn{3}{|c|}{10\%}\\
			\cline{2-4} \cline{5-7}
			\cline{2-4} \cline{5-7}
			&{No-D}&{BM3D}&{DM}&{No-D}&{BM3D}&{DM}\\
			\cline{2-4}\cline{5-7}
			& \multicolumn{6}{c|}{PSNR (dB)/SSIM}\\
			\hline\hline
			ReconNet\cite{kulkarni2016reconnet}&29.68/0.8713&{\bf29.89}/0.8802&29.77/{\bf0.8809}&24.33/0.7232 &23.69/0.6483&{\bf24.81}/{\bf0.7477}\\
			LDIT\cite{metzler2017learned}&32.95/0.9249&32.99/0.9254&{\bf33.68}/{\bf0.9343}&25.86/0.7791&25.95/0.7872&{\bf26.83}/{\bf0.8122}\\
			LDAMP\cite{metzler2017learned}&32.45/0.9186&{\bf32.50}/{\bf0.9198}&32.28/0.9168&25.34/0.7604&25.44/0.7709&{\bf26.03}/{\bf0.7887}\\
			DPDNN\cite{dong2018denoising}&32.61/0.9220&32.60/0.9220&{\bf33.97}/{\bf0.9378}&25.30/0.7668&25.39/0.7762&{\bf27.32}/{\bf0.8262}\\
			NN\cite{gilton2019neumann}&27.78/0.8147&27.29/0.8153&{\bf29.96}/{\bf0.8763}&24.08/0.7025&24.12/0.7102&{\bf24.44}/{\bf0.7312}\\
			${\text{ISTA-Ne}}{{\text{t}}^+}$\cite{zhang2018ista}&33.04/0.9264&33.10/0.9270&{\bf34.40}/{\bf0.9404}&25.56/0.7689&25.66/0.7780&{\bf26.78}/{\bf0.8102}\\
			AMP-Net-6&33.34/0.9292&33.33/0.9261&{\bf34.01}/{\bf0.9370}&26.09/0.7878&26.22/0.7963&{\bf27.05}/{\bf0.8180}\\
			AMP-Net-9&33.65/0.9331&33.59/0.9288&{\bf34.41}/{\bf0.9407}&26.44/0.7975&26.55/0.8045&{\bf27.35}/{\bf0.8262}\\
			\hline
			
		\end{tabular}
		\label{tab:deblock}
	\end{center}
	\vspace{-3mm}
\end{table*}

\subsection{Validating the unfolding strategy of AMP-Net}
\label{Comparison}
{In this subsection, we verify the effectiveness of the unfolding strategy of AMP-Net.
We compare AMP-Net-$K$ with other eight state-of-the-art methods, namely TVAL3~\cite{li2013efficient}, D-AMP~\cite{metzler2016denoising}, ReconNet~\cite{kulkarni2016reconnet}, LDIT~\cite{metzler2017learned}, LDAMP~\cite{metzler2017learned}, DPDNN~\cite{dong2018denoising}, NN~\cite{gilton2019neumann} and ${\text{ISTA-Ne}}{{\text{t}}^+ }$~\cite{zhang2018ista}. In this subsection, all methods reconstruct images block-by-block and do not apply deblocking operations and sampling mask training strategies.
Because ${\text{CSNe}}{{\text{t}}^+}$ reconstructs and deblocks images at the same time and the sampling matrix optimization is important to ${\text{CSNe}}{{\text{t}}^+}$ and DCS, we compare AMP-Net with ${\text{CSNe}}{{\text{t}}^+}$ and DCS in Section \ref{sec: B_M}.
All trainable models are trained on training set 2.}

Table \ref{tab:set11} shows the test results of AMP-Net-6, AMP-Net-9 and other methods on Set11 {at} different CS ratios of 50\%, 40\%, 30\%, 25\%, 10\%, 4\% and 1\%. Table \ref{tab:set11} contains the average PSNR (dB) and SSIM where the best is marked in bold. {And except for ${\text{ISTA-Ne}}{{\text{t}}^+ }$ and AMP-Net-9, which has 9 reconstruction modules, all other deep unfolding models have only 6 reconstruction modules.}

From Table \ref{tab:set11}, it can be found that TVAL3 and D-AMP have worse performance {at} low CS ratios of 4\% and 1\%.
ReconNet does not work well {at} high CS ratios but {at} low CS ratios.
{For {the} block-based image CS problem, LDIT has better performance than LDAMP in our paper which may be because the learned Onsager correction term losses its advantages while the CNN is not deep enough.}
NN shows worse reconstruction results {at} high CS ratios of 50\%, 40\%, 30\% and 25\% but better results than two {model-based} methods {at} 10\%, 4\% {and} 1\%.
And as a deep unfolding method for common image inverse problem, DPDNN does not has a significant advantage in the image CS.
{AMP-Net-6 performs better than DPDNN and NN.} {The reason why AMP-Net is superior to DPDNN may be that AMP-Net is established by analyzing noise items which are highly correlated with the image CS problem, and this strategy makes AMP-Net more specific to the image CS problem.}
{As a model derived from the AMP algorithm, AMP-Net-6 has higher PSNR and SSIM than other AMP-based methods, namely D-AMP, LDIT and LDAMP, {at} most CS ratios.
With the same number of reconstruction modules as 9, ${\text{ISTA-Ne}}{{\text{t}}^ + }$ performs better {at} high CS ratios of 50\%, 40\%, 30\%, 25\%, but AMP-Net has better performance {at} low ratios.}

\begin{figure}
	\vspace{-0.3cm}
	\setlength{\abovecaptionskip}{-5pt}
	\setlength{\belowcaptionskip}{-5pt} 
	\begin{center}
		\includegraphics{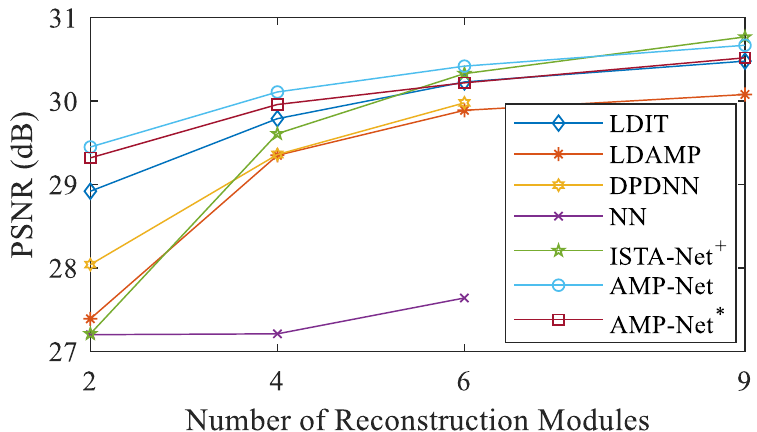}
	\end{center}

	\caption{PSNR of models with different reconstruction numbers on the test set of BSDS500 {at the} CS ratio of 30\%.}
	
	\label{fig:unfolding_30}
\end{figure}

\begin{figure}
	\vspace{-0.3cm}
	\setlength{\belowcaptionskip}{-5pt}
	\setlength{\abovecaptionskip}{-5pt}
	\begin{center}
		\includegraphics{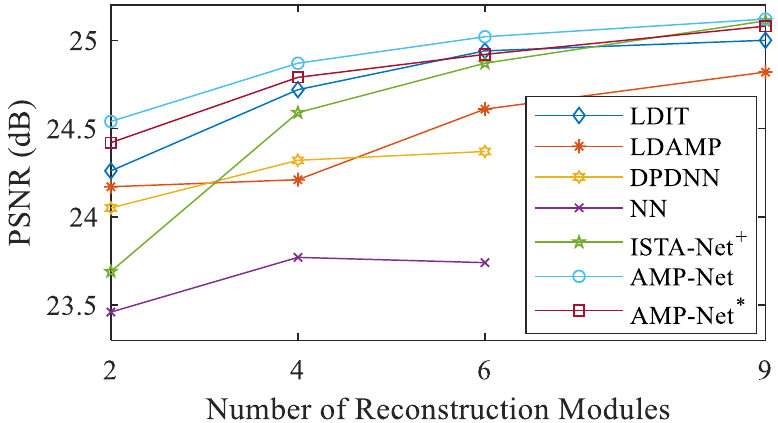}
	\end{center}
	\caption{PSNR of models with different reconstruction numbers on the test set of BSDS500 {at the} CS ratio of 10\%.}
	\label{fig:unfolding_10}
	\vspace{-0.4cm}
\end{figure}

To further evaluate the generalization performance of AMP-Net, we compare AMP-Net-6, AMP-Net-9 with ReconNet, LDIT, LDAMP, ${\text{ISTA-Ne}}{{\text{t}}^ + }$, DPDNN and NN on the test set of BSDS500. Table \ref{tab:bsd500} shows the average PSNR and SSIM where the best is marked in bold.
{From Table \ref{tab:bsd500}, it is clear that AMP-Net-6 has better performance than ReconNet, LDIT, LDAMP, DPDNN and NN. And AMP-Net-9 performs better than ${\text{ISTA-Ne}}{{\text{t}}^ + }$ {at} CS ratios of 10\%, 4\% and 1\%.
	
Moreover, Fig. \ref{fig:unfolding_30} and Fig. \ref{fig:unfolding_10} show {the} average PSNR of 7 deep unfolding methods with different numbers of reconstruction modules tested on the test set of BSDS500 {at} two different CS ratios of 30\% and 10\%. AMP-Net$^{*}$ is AMP-Net with $\alpha _k$ fixed to 1.
It can be found that with different {numbers} of reconstruction modules, AMP-Net-$K$ performs best {at} the CS ratio of 10\%, and is also very competitive when the CS ratio is 30\% and the number of reconstruction modules is smaller than 9.
Significantly, when $\alpha _k$ equals 1, AMP-Net-$K$ still has better performance than LDIT, which may be because our noise estimation strategy is more effective.
We conclude that the unfolding strategy of AMP-Net is effective and AMP-Net has a good generalization performance.}

\subsection{Validating the Capability of the Deblocking Module}
\label{sec: deblock}

In this subsection, we validate the capability of the deblocking module.
{Firstly, to show the effectiveness of the deblocking module on AMP-Net, we train AMP-Net-6, AMP-Net-6-B, AMP-Net-9 and AMP-Net-9-B on training set 1 {at} CS ratios of 30\% and 10\%, respectively.
Secondly, to prove that the deblocking module can be applied in other deep learning methods, we plug the deblocking module in ReconNet, LDIT, LDAMP, ${\text{ISTA-Ne}}{{\text{t}}^+}$, DPDNN and NN, and {train them on training set 1 by employing concatenation functions}. Except for ${\text{ISTA-Ne}}{{\text{t}}^+ }$, which has 9 reconstruction modules, all other deep unfolding models have only 6 reconstruction modules.
Significantly, the deblocking module is attached to ReconNet and NN, and plugged in each iteration of the other four deep unfolding models. It is worth noting that ReconNet uses MSE as its loss function in this subsection because its adversarial loss is only designed for each block.}

\begin{figure*}
	\setlength{\belowcaptionskip}{0pt}
	\setlength{\abovecaptionskip}{-5pt}
	\begin{center}
		\includegraphics{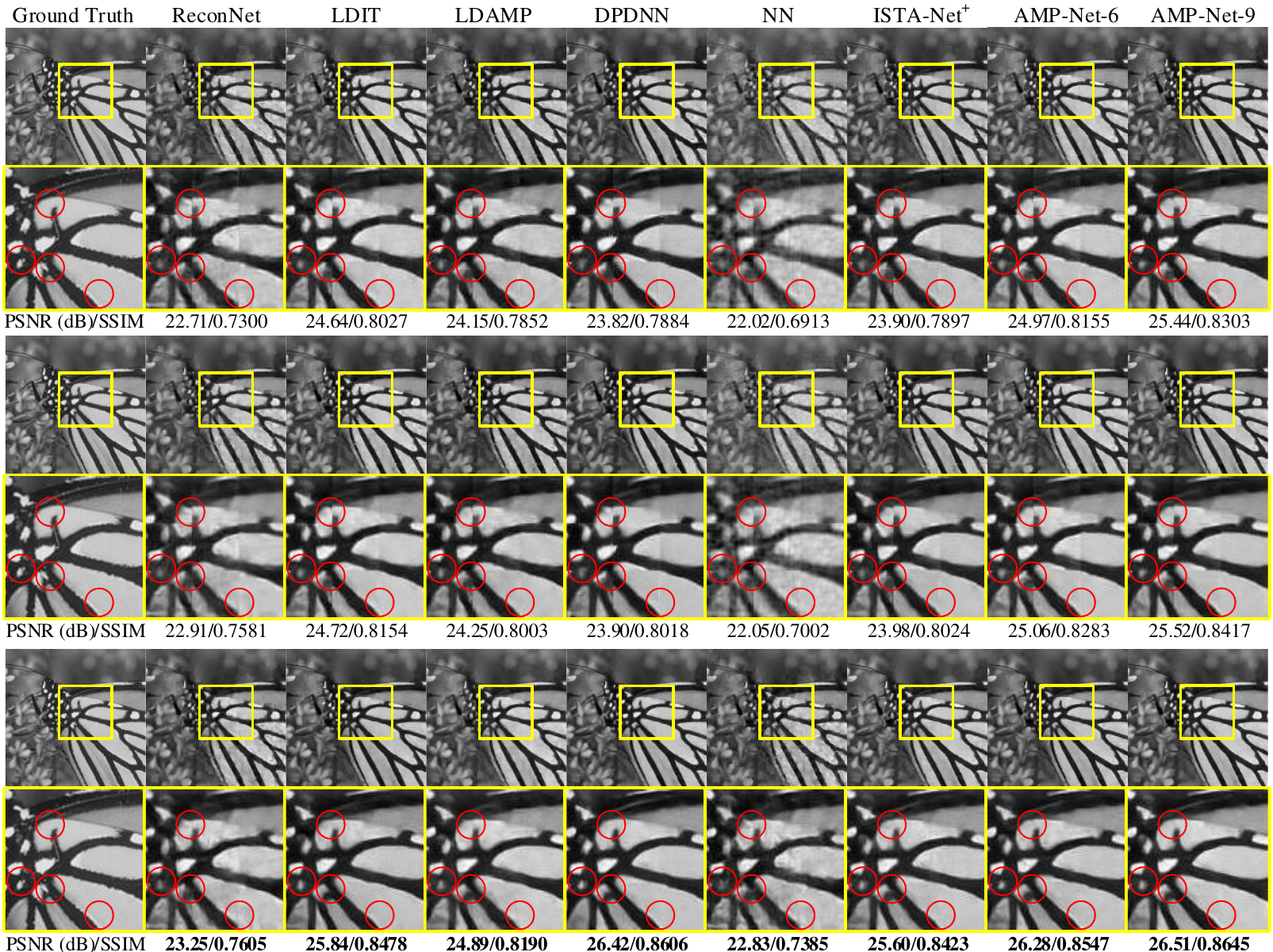}
	\end{center}
	\caption{The reconstruction results of the 8 methods on \emph{Monarch} image {at} the CS ratio of 10\%. Images in the first row, the second row and the third row are reconstructed with no deblocking strategy, BM3D denoiser and the deblocking module, respectively.}
	\label{fig:deblock}
	\vspace{-0.4cm}
\end{figure*}

{Table \ref{tab:deblock} shows the average PSNR and SSIM of models above tested on Set11 {at} CS ratios of 30\% and 10\% under different conditions, including no deblocking, with BM3D, with the deblocking module. BM3D is applied for deblocking using the strategy in \cite{kulkarni2016reconnet}.
Fig. \ref{fig:deblock} shows the reconstruction results of the 8 methods on \emph{Monarch} image in Set11 {at} the CS ratio of 10\% with different deblocking strategies.}

Table \ref{tab:deblock} and Fig. \ref{fig:deblock} are quite revealing in several ways.

\emph{1) The performance improvement with the deblocking module.} {Comparing the PSNR and SSIM in Table \ref{tab:deblock} and Fig. \ref{fig:deblock}, models with the deblocking module have higher PSNR and SSIM than other versions. It benefits from the joint training of deblocking modules and other parameters.
Therefore, we demonstrate that with deblocking modules, reconstruction models can provide better recovery results.}

\emph{2) The analysis of the deblocking ability.} {From Fig. \ref{fig:deblock}, with the same training strategy, models without deblocking modules} still generate images with blocking {artifacts} which {are} clearly visible in the red-circle-marked areas. This means that the blocking {artifacts} are inevitably introduced due to the block-by-block reconstruction of images. {Because BM3D is designed to remove Gaussian noise, the image denoised by BM3D still has visible {artifacts}. However, there is no obvious blocking {artifact} in images generated by models with the deblocking module. Therefore, we demonstrate that with deblocking {modules}, the blocking {artifacts} can be erased effectively.}

\emph{3) {The universality of the deblocking module.}} {From table \ref{tab:deblock}, it is clear that models with the deblocking module outperform models without it. 
It means that the deblocking module can be universal in other deep learning models.}

\subsection{Evaluating the Sampling Matrix Training Strategy}
\label{sec: sampling}
{In this subsection, we validate the improvement of the reconstruction performance on AMP-Net brought by the sampling matrix training strategy. In addition, we numerically show that using the trained sampling matrix of AMP-Net, most methods can have better reconstruction performance.}

{Firstly, we train AMP-Net-6-M and AMP-Net-9-M on the training set 2 {at} CS ratios of 30\%, 10\% and 4\%. Table \ref{tab:sampling_set11} and Table \ref{tab:sampling_bsd500} show the average PSNR and SSIM of AMP-Net-6, AMP-Net-6-M, AMP-Net-9 and AMP-Net-9-M tested on Set11 and the test set of BSDS500. Obviously, with the trained sampling matrix, AMP-Net-6-M and AMP-Net-9-M have higher PSNR and SSIM than AMP-Net-6 and AMP-Net-9 {at} all CS ratios. We emphasize that with the same CS ratio, by training the sampling matrix, PSNR and SSIM can be roughly improved by 2dB and 0.1, respectively.
Therefore, it demonstrates that the sampling matrix training strategy do improve the reconstruction performance of AMP-Net.}

\begin{table}
	\setlength{\abovecaptionskip}{-2pt}
	\begin{center}
		\footnotesize
		\caption{Test results of AMP-Net-6, AMP-Net-6-M, AMP-Net-9 and AMP-Net-9-M on Set11.}
		\begin{tabular}{|l|c|c|c|}
			\hline
			\multirow{2}*{{\bf Method}}&30\%&{10\%}&{4\%}\\
			\cline{2-4}
			& \multicolumn{3}{c|}{PSNR (dB)/SSIM}\\
			\hline\hline
			AMP-Net-6&32.96/0.9254&25.72/0.7750&21.52/0.6100\\
			AMP-Net-6-M&35.56/0.9557&28.74/0.8659&24.53/0.7428\\
			AMP-Net-9&33.40/0.9307&25.93/0.7828&21.69/0.6224\\
			AMP-Net-9-M&35.68/0.9567&28.84/0.8684&24.65/0.7516\\
			\hline
		\end{tabular}
		\label{tab:sampling_set11}
	\end{center}
	\vspace{-0.5cm}
\end{table}

\begin{table}
	\setlength{\belowcaptionskip}{0pt}
	\setlength{\abovecaptionskip}{-2pt}
	\begin{center}
		\footnotesize
		\caption{Test results of AMP-Net-6, AMP-Net-6-M, AMP-Net-9 and AMP-Net-9-M on BSDS500.}
		\begin{tabular}{|l|c|c|c|}
			\hline
			\multirow{2}*{{\bf Method}}&30\%&{10\%}&{4\%}\\
			\cline{2-4}
			& \multicolumn{3}{c|}{PSNR (dB)/SSIM}\\
			\hline\hline
			AMP-Net-6&30.43/0.8840&25.02/0.7095&22.09/0.5779\\
			AMP-Net-6-M&32.91/0.9330&27.48/0.8045&24.68/0.6814\\
			AMP-Net-9&30.67/0.8887&25.12/0.7153&22.22/0.5851\\
			AMP-Net-9-M&32.99/0.9342&27.51/0.8053&24.66/0.6833\\
			\hline
		\end{tabular}
		\label{tab:sampling_bsd500}
	\end{center}
	\vspace{-0.5cm}
\end{table}

\begin{figure*}
	\setlength{\belowcaptionskip}{0pt}
	\setlength{\abovecaptionskip}{-5pt}
	\begin{center}
		\includegraphics{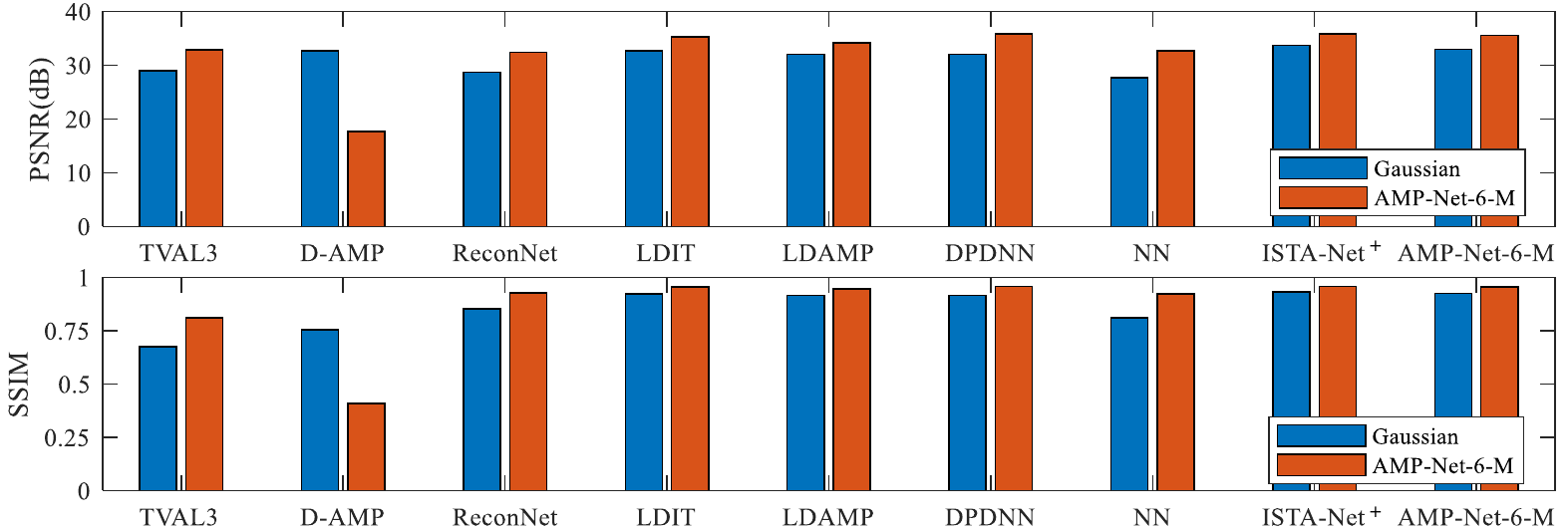}
	\end{center}
	\caption{Test results of different methods with two different sampling matrices tested on Set11 {at} the CS ratio of 30\%. }
	\label{fig:sampling}
	\vspace{-3mm}
\end{figure*}

\begin{table*}
	\setlength{\belowcaptionskip}{0pt}
	\setlength{\abovecaptionskip}{2pt}
	\begin{center}
		\footnotesize
		\caption{The test results of ten comparison methods and AMP-Net-2-BM, AMP-Net-4-BM, AMP-Net-6-BM and AMP-Net-9-BM  on Set11.}
		\begin{tabular}{|l|c|c|c|c|c|c|c|c|}
			\hline
			\multirow{2}*{{\bf Method}} & 50$\%$ & 40$\%$ &30$\%$ &25$\%$ &10$\%$ &4$\%$ &1$\%$ &{\bf Time (s)}\\
			\cline{2-8}
			& \multicolumn{7}{c|}{PSNR (dB)/SSIM}&CPU/GPU \\
			\hline\hline
			
			TVAL3~\cite{li2013efficient}&33.39/0.8157&31.21/0.7531&29.00/0.6764 & 27.63/0.6238 &22.45/0.3758&17.88/0.1997&14.90/0.0646 &2.379/——\\ 
			
			D-AMP~\cite{metzler2016denoising} & 37.34/0.8504  &35.22/0.8078 & 32.64/0.7544 & 31.62/0.7233 & 19.87/0.3757 & 11.28/0.0971 & 5.58/0.0034&39.139/——\\ 
			
			DCS~\cite{Wu2019deep} & 22.30/0.5452  &21.99/0.5033 & 21.98/0.5358 & 21.85/0.5116 & 21.53/0.4546 & 18.03/0.2202 & 17.12/0.3251&1.829/0.036\\ 
			
			ReconNet~\cite{kulkarni2016reconnet} & 37.42/0.9609 &35.48/0.9676 & 33.17/0.9380 & 32.07/0.9246 & 27.63/0.8487 & 24.29/0.7382 & 20.16/0.5431&{\color{red}0.087}/{\color{red}0.004}\\
			${\text{CSNe}}{{\text{t}^+}}$~\cite{shi2019image} & 38.19/0.9739 &36.15/0.9625 & 33.90/0.9449 & 32.76/0.9322 & 27.76/0.8513 & 24.24/0.7412 & 20.09/0.5334&0.448/{\color{green}0.007}\\
			LDIT\cite{metzler2017learned} & 37.06/0.9626 &34.92/0.9470 & 32.69/0.9223 & 31.35/0.9042 & 25.56/0.7691 & 21.45/0.6075 & 17.58/0.4449&{\color{green}0.150}/{\color{blue}0.008}\\
			LDAMP\cite{metzler2017learned} & 35.90/0.9531 &34.07/0.9383 & 32.01/0.9144 & 29.93/0.8783 & 24.94/0.7483 & 21.30/0.5985 & 17.51/0.4409&0.312/0.015\\
			DPDNN~\cite{dong2018denoising}&35.85/0.9532&34.30/0.9411&32.06/0.9145&30.63/0.8924&24.53/0.7392&21.11/0.6029&17.59/0.4459&2.439/0.058\\
			NN~\cite{gilton2019neumann}&31.41/0.8871&29.51/0.8523&27.64/0.8095&26.57/0.7842&22.99/0.6591&20.65/0.5525&17.67/0.4324&10.253/0.058\\
			${\text{ISTA-Ne}}{{\text{t}}^+}$~\cite{zhang2018ista} & 38.08/0.9680 &35.93/0.9537 & 33.66/0.9330 & 32.27/0.9167 & 25.93/0.7840  & 21.14/0.5947 & 17.48/0.4403&1.126/0.027\\
			AMP-Net-2-BM & {39.48}/{0.9781} &{37.52}/{0.9686}  & {35.21}/{0.9530} & {33.92}/{0.9417} & {28.67}/{0.8654} & {24.72}/{0.7562} & {\color{blue}20.41}/{0.5539}&{\color{blue}0.170}/{\color{blue}0.008}\\
			AMP-Net-4-BM & {\color{blue} 40.07}/{\color{blue} 0.9795} &{\color{blue} 38.03}/{\color{blue} 0.9705} & {\color{blue} 35.67}/{\color{blue} 0.9564} & {\color{blue} 34.38}/{\color{blue} 0.9451} & {\color{blue} 29.05}/{\color{blue} 0.8728} & {\color{blue} 25.07}/{\color{blue} 0.7680} & {20.35}/{\color{blue} 0.5563}&{0.333}/0.014\\
			AMP-Net-6-BM & {\color{green} 40.27}/{\color{green} 0.9804} &{\color{green} 38.23}/{\color{green} 0.9713} & {\color{green} 35.90}/{\color{green} 0.9574} & {\color{green} 34.59}/{\color{green} 0.9477} & {\color{red} 29.45}/{\color{red} 0.8787} & {\color{green} 25.16}/{\color{green} 0.7692} & {\color{red} 20.57}/{\color{red} 0.5639}&0.500/0.020\\
			AMP-Net-9-BM & {\color{red} 40.34}/{\color{red} 0.9807} &{\color{red} 38.28}/{\color{red} 0.9715} & {\color{red} 36.03}/{\color{red} 0.9586} & {\color{red} 34.63}/{\color{red} 0.9481} & {\color{green} 29.40}/{\color{green} 0.8779} & {\color{red} 25.26}/{\color{red} 0.7722} & {\color{green} 20.20}/{\color{green} 0.5581}&0.544/0.027\\
			\hline
		\end{tabular}
		\label{tab:set11_BM}
	\end{center}
	\vspace{-3mm}
\end{table*}

{Secondly, we fix and use the sampling matrix of AMP-Net-6-M to train or test other models, including TVAL3, DAMP, ReconNet, LDIT, LDAMP, DPDNN, NN and ${\text{ISTA-Ne}}{{\text{t}}^+ }$. All trainable models are trained on the training set 2. Except for ${\text{ISTA-Ne}}{{\text{t}}^+ }$, which has 9 reconstruction modules, all other deep unfolding models have only 6 reconstruction modules.
Fig. \ref{fig:sampling} shows the histogram of the average PSNR and SSIM of models with the random sampling matrix in Gaussian distribution and models with the sampling matrix of AMP-Net-6-M {at} the CS ratio of 30\%. 
Due to characteristics of the Monte-Carlo approximation of AMP, the sampling matrix of D-AMP should be i.i.d. Gaussian matrix~\cite{8713501}. That might be the reason why the {performance} of D-AMP {decreases} using the sampling matrix of AMP-Net-6-M. However, it is obvious that with the sampling matrix of AMP-Net-6-M, most models can perform better. }
 
\subsection{Validating the performance of AMP-Net-$K$-BM}
\label{sec: B_M}
{In this subsection, we evaluate the performance of AMP-Net-$K$-BM, and compare it with other state-of-the-art methods. To do so, we train AMP-Net-$K$-BM, where $K$ contains 2, 4, 6 and 9 in this subsection.
Methods for comparison include TVAL3, D-AMP, DCS, ReconNet, ${\text{CSNe}}{{\text{t}}^+}$, LDIT, LDAMP, DPDNN, NN and ${\text{ISTA-Ne}}{{\text{t}}^+}$.
{In this subsection, comparison models are trained and tested based on the settings in their original papers. In detail, the sampling matrices of DCS, ReconNet and ${\text{CSNe}}{{\text{t}}^+}$ are trained. ReconNet combines the adversarial loss and MSE for training, and enables BM3D for deblocking.  AMP-Net-$K$-BM and ${\text{CSNe}}{{\text{t}}^+}$ are trained on training set 1 because of the existence of trainable deblocking operations, while the other models take training set 2 due to their characteristics of image recovery in a block-by-block way.}
Moreover, except for ${\text{ISTA-Ne}}{{\text{t}}^+ }$ with 9 reconstruction modules, the rest compared deep unfolding models have only 6 reconstruction modules.

\begin{table*}
	\setlength{\belowcaptionskip}{0pt}
	\setlength{\abovecaptionskip}{2pt}
	\begin{center}
		\footnotesize
		\caption{The test results of two classical deep learning methods, five deep unfolding methods, AMP-Net-2-BM, AMP-Net-4-BM, AMP-Net-6-BM and AMP-Net-9-BM on the test set of BSDS500. PN refers to the parameter number of each model. }
		\begin{tabular}{|l|c|c|c|c|c|c|c|c|}
			\hline
			\multirow{2}*{{\bf Method}} & 50$\%$ & 40$\%$ &30$\%$ &25$\%$ &10$\%$ &4$\%$ &1$\%$ &\multirow{2}*{\bf PN} \\
			\cline{2-8}
			& \multicolumn{7}{c|}{PSNR (dB)/SSIM}& \\
			\hline\hline
			
			ReconNet~\cite{kulkarni2016reconnet} & 35.32/0.9631 &33.46/0.9458 & 31.49/0.9171 & 30.51/0.8971 & 26.85/0.7837 & 24.44/0.6693 & 21.45/0.5322&{\color{red} 22914}\\
			${\text{CSNe}}{{\text{t}^+}}$~\cite{shi2019image} & 35.89/0.9677 &33.96/0.9513 & 31.94/0.9251 & 30.91/0.9067 & 27.01/0.7949 & 24.41/0.6747 & 21.42/0.5261&370560\\
			LDIT\cite{metzler2017learned} & 34.27/0.9453 &32.23/0.9184 & 30.23/0.8799 & 29.10/0.8523 & 24.94/0.7040 & 22.04/0.5759 & 19.00/0.4564&{\color{blue} 114624}\\
			LDAMP\cite{metzler2017learned} & 33.45/0.9359 &31.79/0.9116 & 29.89/0.8724 & 28.35/0.8297 & 24.61/0.6920 & 21.93/0.5721 & 18.94/0.4512&229248\\
			DPDNN~\cite{dong2018denoising}&33.56/0.9373&32.05/0.9164&29.98/0.8759&28.87/0.8491&24.37/0.6863&21.80/0.5716&18.97/0.4544&1363712\\
			NN~\cite{gilton2019neumann}&30.47/0.8882&28.84/0.8511&27.23/0.8037&26.42/0.7757&23.44/0.6443&21.49/0.5451&19.06/0.4474&2954516\\
			${\text{ISTA-Ne}}{{\text{t}}^+}$~\cite{zhang2018ista} & 34.92/0.9510 &32.87/0.9264 & 30.77/0.8901 & 29.64/0.8638 & 25.11/0.7124  & 21.82/0.5661 & 18.92/0.4529&336978\\
			AMP-Net-2-BM & {36.89}/{0.9720} &{34.88}/{0.9567}  & {32.74}/{0.9319} & {31.62}/{0.9134} & {27.50}/{0.8051} & {24.77}/{0.6873} & {21.77}/{0.5416}&{\color{green} 76418}\\
			AMP-Net-4-BM & {\color{blue} 37.30}/{\color{blue} 0.9735} &{\color{blue} 35.22}/{\color{blue} 0.9587} & {\color{blue} 33.04}/{\color{blue} 0.9348} & {\color{blue} 31.88}/{\color{blue} 0.9168} & {\color{blue} 27.70}/{\color{blue} 0.8108} & {\color{blue} 24.92}/{\color{blue} 0.6938} & {\color{blue} 21.79}/{\color{blue} 0.5473}&{152836}\\
			AMP-Net-6-BM & {\color{green} 37.48}/{\color{green} 0.9744} &{\color{green} 35.34}/{\color{green} 0.9594} & {\color{green} 33.17}/{\color{green} 0.9358} & {\color{green} 32.01}/{\color{green} 0.9188} & {\color{green} 27.82}/{\color{green} 0.8133} & {\color{green} 24.95}/{\color{green} 0.6949} & {\color{red} 21.90}/{\color{green} 0.5501}&229254\\
			AMP-Net-9-BM & {\color{red} 37.51}/{\color{red} 0.9750} &{\color{red} 35.43}/{\color{red} 0.9600} & {\color{red} 33.24}/{\color{red} 0.9367} & {\color{red} 32.05}/{\color{red} 0.9195} & {\color{red} 27.84}/{\color{red} 0.8138} & {\color{red} 25.04}/{\color{red} 0.6971} & {\color{green} 21.82}/{\color{red} 0.5503}&343881\\
			\hline
		\end{tabular}
		\label{tab:bsd500_BM}
	\end{center}
	\vspace{-3mm}
\end{table*}

\begin{table*}
	\setlength{\belowcaptionskip}{0pt}
	\setlength{\abovecaptionskip}{2pt}
	\begin{center}
		\footnotesize
		\caption{The test results of AMP-Net-$K$, AMP-Net-$K$-B, AMP-Net-$K$-M and AMP-Net-$K$-BM on the test set of BSDS500.}
		\begin{tabular}{|c|c|c|c|c|c|c|c|c|}
			\hline
			
			\multirow{3}*{{\bf CS ratio}}&AMP-Net&AMP-Net&AMP-Net&AMP-Net&AMP-Net&AMP-Net&AMP-Net&AMP-Net\\
			&-6&-6-B&-6-M&-6-BM&-9&-9-B&-9-M&-9-BM\\
			\cline{2-5}\cline{6-9}
			& \multicolumn{8}{c|}{PSNR (dB)/SSIM}\\
			\hline\hline
			30\%&30.43/0.8840&31.07/0.8956&32.91/0.9330&33.17/0.9358 &30.67/0.8887&31.30/0.8995&32.99/0.9342&33.24/0.9367\\
			
			10\%&25.02/0.7095&25.72/0.7376&27.48/0.8045&27.82/0.8133&25.12/0.7153&25.84/0.7424&27.51/0.8053&27.84/0.8138\\
			
			4\%&22.09/0.5779&22.86/0.6169&24.68/0.6814&24.95/0.6949&22.22/0.5851&22.95/0.6218&24.66/0.6833&25.04/0.6971\\
			\hline
			
		\end{tabular}
		\label{tab:AMP_net_all}
	\end{center}
	\vspace{-0.6cm}
\end{table*}

\begin{table}
	\setlength{\belowcaptionskip}{0pt}
	\setlength{\abovecaptionskip}{2pt}
	\begin{center}
		\footnotesize
		\caption{The time consuming analysis (GPU/CPU) and parameter number (PN) of each reconstruction module of different versions of AMP-Net.}
		\begin{tabular}{|c|c|c|c|c|}
			\hline
			\multirow{2}*{{\bf Index}}&AMP-Net&AMP-Net&AMP-Net&AMP-Net\\
			&-$K$&-$K$-B&-$K$-M&-$K$-BM\\
			\hline\hline
			{\bf Times (s)}&{0.033}/0.002&{0.060}/0.003&{0.033}/0.002&{0.060}/0.003 \\
			{\bf PN}&19105&38209&19105&38209 \\
			\hline
			
		\end{tabular}
		\label{tab:AMP_net_all_Time_PN}
	\end{center}
	\vspace{-0.6cm}
\end{table}

Table \ref{tab:set11_BM} and Table \ref{tab:bsd500_BM} show the test results on Set11 and the test set of BSDS500 {at} different CS ratios, where the best, the second and the third results are marked with red front, green front and {blue} front, respectively. The last column of Table {\ref{tab:set11_BM}} is the time consuming analysis (average reconstruction time of a 256$\times$256 gray-scale image) of each method, in which the time analysis of ReconNet does not include the deblocking of BM3D. The last column of Table \ref{tab:bsd500_BM} {contains} the parameter number of each deep-learning-based model.
Fig. \ref{fig:Monarch} and Fig. \ref{fig:Parrots} show the reconstructed \emph{Monarch} and \emph{Parrots} images in Set11 at CS ratios of 30\% and 10\%.

It can be noticed that with deblocking modules and the sampling matrix training strategy, AMP-Net outperfroms other methods. Especially, compared with ReconNet and ${\text{CSNe}}{{\text{t}}^+}$, which have both deblocking operations and trained sampling matrices, AMP-Net-$K$-BM still works better due to its deep unfolding strategy. And as the number of the reconstruction modules increases, the performance gets better.
In addition, from the last columns of Table \ref{tab:set11_BM} and \ref{tab:bsd500_BM},
it can be noticed that when $K$ is small, AMP-Net-$K$-BM reconstructs image fast and has a small parameter number, while maintaining a good performance.}

{In addition, in order to visually demonstrate the improvement of the performance of AMP-Net brought by the deblocking module and {the} sampling matrix training strategy, we compare AMP-Net-$K$, AMP-Net-$K$-B, AMP-Net-$K$-M and AMP-Net-$K$-BM. Table \ref{tab:AMP_net_all} shows the average PSNR and SSIM of different versions of AMP-Net with different numbers of reconstruction modules {at} CS ratios of 30\%, 10\% and 4\%. And Table \ref{tab:AMP_net_all_Time_PN} contains the time consuming analysis (average reconstruction time  (GPU/CPU) of a 256$\times$256 gray-scale image) and the parameter number of each reconstruction module of different versions of AMP-Net. Obviously, the  deblocking module and trained sampling matrix can improve the performance of AMP-Net, and AMP-Net with both of them performs best. Meanwhile, it can be noticed that combined with the deblocking module, the model can have a larger number of parameters but better performance.}

\section{Conclusion}
\label{conclusion}
In this paper, we design a deep unfolding model named
AMP-Net based on the denoising perspective of the AMP
algorithm to solve the visual image CS problem.
{Experimental results {show} the effectiveness of the unfolding, the deblocking and the sampling matrix training strategy of AMP-Net. Meanwhile, generalized analysis of the deblocking and the sampling matrix training strategy over other reconstruction methods is also conducted. It reveals the versatility of these two strategies in other reconstruction methods. Finally, the performance of AMP-Net-$K$-BM is validated with a comprehensive comparison with {state-of-the-art} reconstruction methods. Results demonstrate that AMP-Net-2-BM has better performance than other ten state-of-the-art methods. And as the number of the reconstruction module increases, the results get better.}

 \begin{figure*}
	\setlength{\belowcaptionskip}{0pt}
	
	\setlength{\abovecaptionskip}{-5pt}
	\begin{center}
		\includegraphics{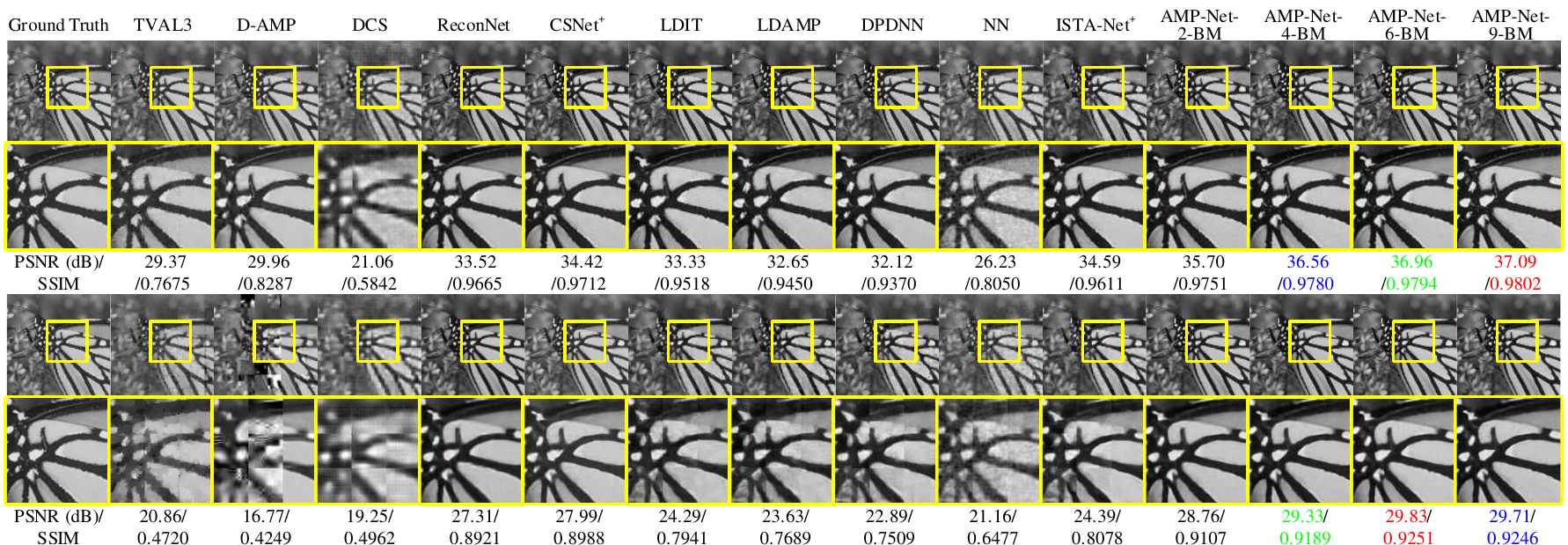}
	\end{center}
	\caption{The reconstruction results on \emph{Monarch} image in Set11 {at} CS ratios of 30\% and 10\%. The first row is images reconstructed {at the} CS ratio of 30\% and the second row is images reconstructed {at the} CS ratio of 10\%.}
	\label{fig:Monarch}
	\vspace{-3mm}
\end{figure*}
\begin{figure*}
	\setlength{\belowcaptionskip}{0pt}
	\setlength{\abovecaptionskip}{-5pt}
	\begin{center}
		\includegraphics{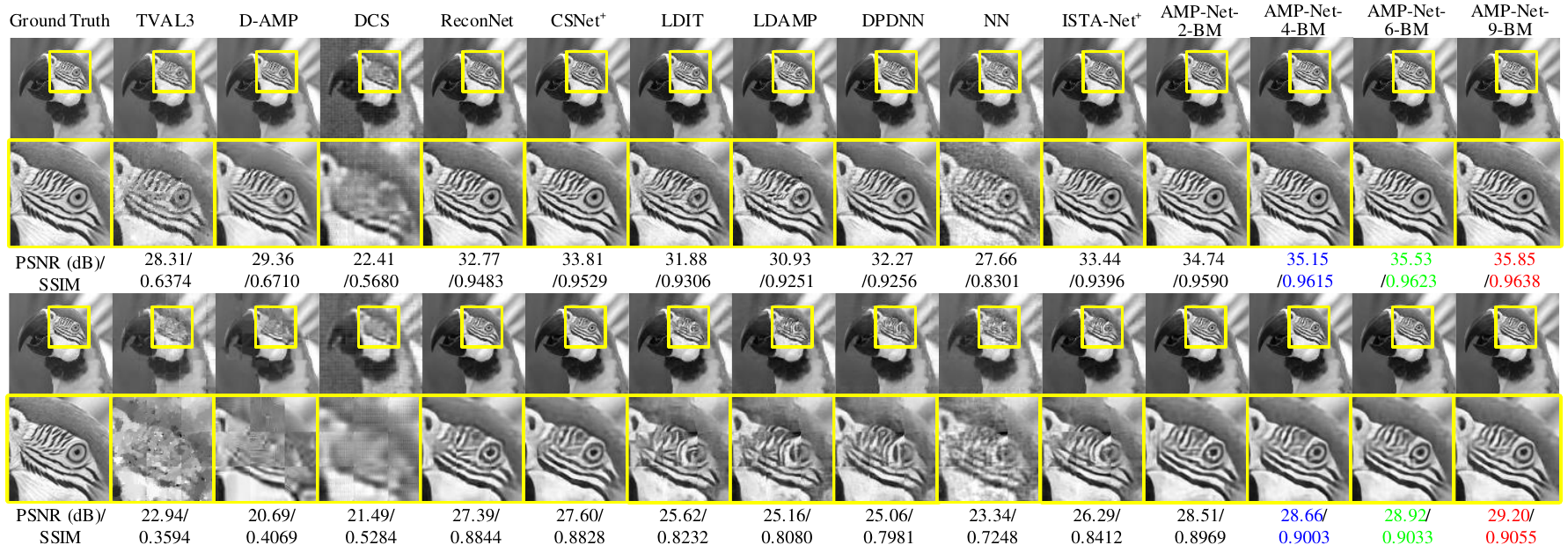}
	\end{center}
	\caption{The reconstruction results on \emph{Parrots} image in Set11 {at} ratios of 30\% and 10\%. The first row is images reconstructed {at the} CS ratio of 30\% and the second row is images reconstructed {at the} CS ratio of 10\%.}
	\label{fig:Parrots}
	\vspace{-3mm}
\end{figure*}


\appendices

\section{Derivation of (\ref{equal5})}

\label{derivation1}
We have $\mathbf{y} = \mathbf{Ax}$ and ${\mathbf{z}^k}=\mathbf{y} - \mathbf{A}{\mathbf{x}^k}$. By combining two of them and the linear operation in (\ref{equal4}), we can get
\allowdisplaybreaks[4]
\begin{align}
\label{equal18}
{\mathbf{A}^{\operatorname{T}}}&{\mathbf{z}^0} + {\mathbf{x}^0} = {\mathbf{A}^{\operatorname{T}}}(\mathbf{y} - \mathbf{A}{\mathbf{x}^0}) + {\mathbf{x}^0} 
= {\mathbf{A}^{\operatorname{T}}}(\mathbf{A}{\bar{\mathbf{x}}} - \mathbf{A}{\mathbf{x}^0}) + {\mathbf{x}^0} \notag \\
&= {\mathbf{A}^{\operatorname{T}}}\mathbf{A}(\bar{\mathbf{x}} - {\mathbf{x}^0}) + {\mathbf{x}^0} 
= {\mathbf{A}^{\operatorname{T}}}\mathbf{A}(\bar{\mathbf{x}} - {\mathbf{x}^0}) - (\bar{\mathbf{x}} - \bar{\mathbf{x}^0}) + \bar{\mathbf{x}} \notag \\
&= \bar{\mathbf{x}} + ({\mathbf{A}^{\operatorname{T}}}\mathbf{A} - \mathbf{I})  (\bar{\mathbf{x}} - {\mathbf{x}^0}).
\end{align}

\section{The Gradient of $\mathbf{A}$}
\label{derivation3}
We present the gradient of $\mathbf{A}$ in the backpropagation training process. Because the sampling matrix is used for processing image blocks in sampling and reconstruction, we only consider the one-block case. In addition, because $\mathfrak{S}( \cdot ,n)$, ${\mathfrak{S}^{ - 1}}( \cdot ,n)$, $\operatorname{vec}( \cdot )$  and ${\operatorname{vec}^{ - 1}}( \cdot )$ do not introduce floating-point calculations, the gradients calculations can exclude them. Define the gradient of $\mathbf{A}$ as ${\nabla _{\mathbf{A}}}\mathfrak{L} = {\mathbf{A}_{{\text{Sam}}}} + {\mathbf{A}_{\text{Rec} }}$, where ${\mathbf{A}_{{\text{Sam}}}}$ is the gradient in the sampling model and ${\mathbf{A}_{{\text{Rec}}}}$ denotes the gradient in the reconstruction model.

Assume that the gradient of ${\mathbf{x}_{i}^0}$ is known and is expressed as ${\nabla _{\mathbf{x}_{i}^0}}\mathfrak{L}$, then according to (\ref{equal8}) and (\ref{equal9}),  ${\mathbf{A}_{{\text{Sam}}}}$ can be written as
\begin{align}
{\mathbf{A}_{{\text{Sam}}}} = {\mathbf{B}^{\operatorname{T}}}  {\nabla _{\mathbf{x}_{i}^0}}\mathfrak{L}  \bar{\mathbf{x}}_{i}^{\operatorname{T}}.
\end{align}
${\mathbf{A}_{{\text{Rec}}}}$ can be expressed as 
\begin{align}
{\mathbf{A}_{{\text{Rec}}}} = \sum\limits_{k = 1}^K {\mathbf{A}_{\text{Rec}}^k},
\end{align}
where $\mathbf{A}_{\text{Rec}}^k$ is the gradient in $k$-th reconstruction module and $K$ is the number of reconstruction modules. Assume that the gradient of {$\mathbf{x}_{i}^{k}$} is known and is expressed as { ${\nabla _{\mathbf{x}_{i}^{k}}}\mathfrak{L}$}, then according to (\ref{equal12}), $\mathbf{A}_{\text{Rec}}^k$ can be expressed as
\begin{align}
\mathbf{A}_{\text{Rec} }^k = {\alpha _k}\mathbf{A}({\nabla _{\mathbf{x}_{i}^{k}}}\mathfrak{L})^{\operatorname{T}}&\hat{\mathbf{x}}_{i}^{k-1}+{\alpha _k}\mathbf{A}(\hat{\mathbf{x}}_{i}^{k-1})^{\operatorname{T}}{\nabla _{\mathbf{x}_{i}^{k}}}\mathfrak{L},\\
\hat{\mathbf{x}}_{i}^{k-1} = \bar{\mathbf{x}}{_{i}} -&{\mathbf{x}}{_{i}}^{k-1} -\operatorname{vec}({\mathfrak{N}_k}(\mathbf{X}_{i}^{k-1})).
\end{align}
Furthermore, ${\nabla _{\mathbf{x}_{i}^{k-1}}}\mathfrak{L}$ can be calculated from ${\nabla _{\mathbf{x}_{i}^{k}}}\mathfrak{L}$, and it can be expressed as
\begin{align}
{\nabla _{\mathbf{x}_{i}^{k-1}}}\mathfrak{L} &= (\mathbf{I}-{\alpha _k}{\mathbf{A}^{\operatorname{T}}}\mathbf{A} ){\nabla _{\mathbf{x}_{i}^{k}}}\mathfrak{L}  \notag \\
+& {({\nabla _{\mathbf{x}_{i}^{k-1}}}\operatorname{vec}({\mathfrak{N}_k}(\mathbf{X}_{i}^{k-1})))^{\operatorname{T}}}  (\mathbf{I}-{\alpha _k}{\mathbf{A}^{\operatorname{T}}}\mathbf{A}){\nabla _{\mathbf{x}_{i}^{k}}}\mathfrak{L}.
\end{align}

\section{Derivation of (\ref{equal12})}
\label{derivation2}
Different from (\ref{equal18}), we introduce a new parameter $\alpha $, and we can get
\begin{align}
\label{equal26}
&{\alpha\mathbf{A}^{\operatorname{T}}}{\mathbf{z}^0} + {\mathbf{x}^0} = {\alpha\mathbf{A}^{\operatorname{T}}}(\mathbf{y} - \mathbf{A}{\mathbf{x}^0}) + {\mathbf{x}^0} \notag \\
&= {\alpha\mathbf{A}^{\operatorname{T}}}(\mathbf{A}\bar{\mathbf{x}} - \mathbf{A}{\mathbf{x}^0}) + {\mathbf{x}^0} \notag \\
&= {\alpha\mathbf{A}^{\operatorname{T}}}\mathbf{A}(\bar{\mathbf{x}} - {\mathbf{x}^0}) - (\bar{\mathbf{x}} - {\mathbf{x}^0}) + \bar{\mathbf{x}} \notag \\
&= \bar{\mathbf{x}} + ({\alpha\mathbf{A}^{\operatorname{T}}}\mathbf{A} - \mathbf{I})  (\bar{\mathbf{x}} - {\mathbf{x}^0}).
\end{align}
By applying (\ref{equal26}) into the image block process and developing it to the iterative version, we can get
\begin{align}
\mathbf{x}_{i}^{k} = &{\alpha _k}{\mathbf{A}^{\operatorname{T}}}\mathbf{z}_{i}^{k-1} + \mathbf{x}_{i}^{k-1}  \notag
\\&- ({\alpha _k}{\mathbf{A}^{\operatorname{T}}}\mathbf{A} - \mathbf{I})  \operatorname{vec}({\mathfrak{N}_k}(\mathbf{X}_{i}^{k-1})).
\end{align}
The process above is the derivation of (\ref{equal12}).



\ifCLASSOPTIONcaptionsoff
  \newpage
\fi

\end{document}